%% file: ms2.tex
\documentclass[oldversion, letterpaper]{aa}         

\usepackage{aalongtable}
\usepackage{natbib}\usepackage{graphicx}      
\bibpunct{(}{)}{;}{a}{}{,}                

\newcommand{\kms}{km\,s$^{-1}$}     
\newcommand{\sqcm}{cm$^{-2}$}  
\newcommand{\mfu}{M$_\odot$\,yr$^{-1}$}

\newcommand{\lya}{Lyman-$\alpha$}
\newcommand{\hi}{\ion{H}{i}} 
\newcommand{\hw}{\ion{H}{ii}} 
\newcommand{\os}{\ion{O}{vi}} 
\newcommand{\cf}{\ion{C}{iv}}

\newcommand{\cts}{\ion{C}{ii}$^*$} 
\newcommand{\cw}{\ion{C}{ii}} 
\newcommand{\ntot}{74} 
\newcommand{\ndla}{63} 
\newcommand{\nsub}{11} 
\newcommand{\ndlaigm}{58} 

\begin{document}

\title{\cf\ absorption in damped and sub-damped \lya\ systems}
\subtitle{Correlations with metallicity and implications for galactic winds at
  $z\approx2-3$\thanks{Based on observations taken with the Ultraviolet
  and Visual Echelle Spectrograph (UVES) on the Very Large Telescope (VLT)
  Unit 2 (Kueyen) at Paranal, Chile, operated by ESO.}}

\author{Andrew J. Fox\inst{1},  C\'edric Ledoux\inst{2}, Patrick
  Petitjean\inst{1, 3}, \& Raghunathan Srianand\inst{4}}
\institute{Institut d'Astrophysique de Paris, UMR7095 CNRS,
  Universit\'e Pierre et Marie Curie, 98bis Blvd Arago, 75014
  Paris, France \and
  European Southern Observatory, Alonso de C\'ordova 3107, Casilla
  19001, Vitacura, Santiago 19, Chile \and 
  LERMA, Observatoire de Paris,
  61 Avenue de l'Observatoire, 75014 Paris, France \and
  IUCAA, Post Bag 4, Ganesh Khind, Pune 411 007, India}

\date{Received April 13, 2007 / Accepted July 27, 2007}  

\authorrunning{Fox et al.}
\titlerunning{\cf\ in DLAs and sub-DLAs} 

\abstract{
We present a study of \cf\ absorption in a sample of \ndla\ damped 
\lya\ (DLA) systems and \nsub\ sub-DLAs in the redshift range 
$1.75\!<\!z_{\rm abs}\!<\!3.61$, using a dataset of high-resolution 
(6.6\,\kms\ FWHM), high signal-to-noise VLT/UVES spectra.
Narrow and broad \cf\ absorption line components indicate the presence
of both warm, photoionized and hot, collisionally ionized gas.
We report new correlations between the metallicity (measured in the
neutral-phase) and each of the \cf\ column density, the
\cf\ total line width, and the maximum \cf\ velocity.
We explore the effect on these correlations of the sub-DLAs, 
the proximate DLAs (defined as those within 5\,000\,\kms\ of the quasar), 
the saturated absorbers, and
the metal line used to measure the metallicity,
and we find the correlations to be robust.
There is no evidence for any difference between the measured properties of DLA
\cf\ and sub-DLA \cf.
In 25 DLAs and 4 sub-DLAs, covering 2.5\,dex in [Z/H], 
we directly observe \cf\ moving above the escape speed, 
where $v_{\rm esc}$ is derived from the total line width of the
neutral gas profiles. These high-velocity \cf\ clouds, unbound from the central
potential well, can be interpreted
as highly ionized outflowing winds, which are
predicted by numerical simulations of galaxy feedback.
The distribution of \cf\ column density in DLAs and sub-DLAs is similar to
the distribution in Lyman Break galaxies, where winds are
directly observed, supporting the idea that supernova feedback 
creates the ionized gas in DLAs.
The unbound \cf\ absorbers show a median mass flow rate of
$\sim$22\,($r$/40\,kpc)\mfu, where $r$ is the characteristic \cf\ radius. 
Their kinetic energy fluxes are large
enough that a star formation rate (SFR) of $\sim2$\,\mfu\
is required to power them.
}
\maketitle

\section{Introduction}
To study the properties of high-redshift galaxies in a
luminosity-independent manner, one can
analyze the absorption lines imprinted by their gaseous halos on the
spectra of background quasars.
Such halos are thought to give rise to QSO
absorption line system of various \hi\ column densities: 
the damped \lya\ systems (DLAs), with 
log\,$N_{\rm \hi}>20.3$, the sub-DLAs with 
$19.0<{\rm log}\,N_{\rm \hi}<20.3$, and 
the Lyman limit systems (LLSs) with 
$17.0<{\rm log}\,N_{\rm \hi}<19.0$.
Observations have shown that highly ionized gas, as detected in
\os\ and \cf\ absorption, is present in each of these categories of absorber
at $z\ga2$: in DLAs \citep{Lu96, Le98, WP00a, Fo07}, in sub-DLAs, 
\citep{DZ03, Pe03, Ri05}, and in LLSs \citep{Be94, KT97, KT99}. 
As one progresses down in \hi\ column density 
from DLAs to LLSs, one may be
sampling progressively more remote (and more highly ionized)
regions of Galactic halos, with most gas in LLSs lying outside the
halo virial radius \citep{Ma03,Da99}. 
Even some \os\ absorbers associated with \lya\ forest clouds, which
are thought to represent the low-density intergalactic medium (IGM), may
arise in extended galaxy halos or feedback zones from galactic
outflows \citep{Be05, Si06}. 

Studying this protogalactic plasma
allows one to address two key themes of extragalactic astronomy: 
galactic winds and the metal budget.
Galactic winds are common at high redshift \citep{Ve05}, and must be
present in order to enrich the IGM up to its 
observed metallicity \citep{Ag01, Ag05, Ar04}.
Simulations predict that the level of ionization in winds is high
\citep{OD06, KR07, Fa07}, and direct observations of absorption in
high-ionization lines have been made in and around Lyman Break
galaxies \citep[LBGs;][]{Pe00, Pe02, Sh03, Ad05}.
Low-redshift studies of galactic outflows have also found a
high-ionization component \citep{He01, St04}. 
Since DLAs represent the largest reservoirs of neutral gas for
high-redshift star formation \citep{Wo05}, they are natural
sites to look for supernova-driven winds and plasma halos in general.

On the second front, ionized halos are
important because of their potential ability to close the metal budget
at $z\approx2$: there is currently a discrepancy 
between the global density of metals predicted by integrating the star
formation history, and the density of metals actually observed
\citep{Pe99, Fe05, Bo05, Bo06, Bo07, SF07}.
The contribution of plasma halos to the metal budget will be
particularly significant if the plasma is hot and
collisionally ionized, because the cooling times in low-metallicity,
low-density, hot halos are extremely long, and so gas injected into
these environments can become locked up until the current epoch. 
For $10^6$\,K gas at $10^{-3}$\,cm$^{-3}$ and solar metallicity, 
$t_{\rm cool}$ is $2.4\times10^{8}$\,yr \citep{HB90}, 
so assuming that to first order $t_{\rm cool}\propto Z^{-1}$, we find that
the cooling time in gas at one-hundredth solar metallicity (as seen in
DLAs) would be approximately equal to the Hubble time. Finding the gas
in hot halos is therefore important for tracing the history of the
cosmic metals. 

In a recent paper \citep[][ hereafter Paper I]{Fo07}, we discussed the
first observations of \os\ absorption in DLAs, finding evidence for a
hot ionized medium that, modulo certain assumptions on metallicity and
ionization, typically contains $\ga$40\% as many baryons
and metals as there are in the neutral phase. Though 12 DLAs with \os\
detections were found, detailed kinematic measurements of the \os\
absorption are difficult due to the high density of blends with the
\lya\ forest. However, if one instead traces the ionized gas with \cf, whose
lines lie redward of the \lya\ forest, the 
blending problems are avoided. Thus, although \cf\ may trace a lower
temperature phase of plasma than \os, it is a better ion to study for
building a sample of statistical size. Partly for this reason, the 
properties of \cf\ absorption in the IGM at $z>2$ have been studied at
length \citep{CS98, El00, Sy03, Bo03, Ar04, Ag05, So05, So06, Sc06a, Sy07}.

Previous observations of \cf\ absorption in DLAs \citep{Lu96, Le98, WP00a} 
and sub-DLAs \citep{DZ03, Pe03, Ri05, Pe07}
have found the \cf\ profiles generally occupy a more extended (though
overlapping) velocity range than the neutral gas profiles. 
In an attempt to explain these observations, \citet{WP00b} tested a
model of gas falling radially onto centrifugally-supported exponential
disks, and found it was unable to reproduce the observed \cf\ kinematics.
On the other hand, \citet{Ma03} found that a model in which hot gas in
halos and sub-halos gives rise to the \cf\ absorption in DLAs was generally
successful in explaining the kinematics.

We continue the study of \cf\ in DLAs in this paper.
We are partly motivated by the recent work of \citet{Le06}, who
reported a correlation between low-ion line width $\Delta v_{\rm Neut}$ and
metallicity [Z/H] in a sample of 70 DLAs and sub-DLAs, covering
over two orders of magnitude in metallicity 
\citep[see also][]{WP98, Mu07, Pr07a}.
Since the line widths of the neutral species are thought to be
gravitationally-dominated, the broader lines may be tracing the more
massive halos, and so the $\Delta v_{\rm Neut}$-[Z/H] correlation has
been interpreted as evidence for an underlying mass-metallicity relation.
A natural follow-on question is whether a similar correlation
exists between the \emph{high-ion} line width and DLA metallicity. 
In this paper we investigate whether such a correlation exists, as
well as exploring other relations between the properties 
of \cf\ and those of the neutral gas.
To maximize our sample size, we include observations of \cf\
absorption in both DLAs and sub-DLAs (and we compare the properties of
the \cf\ absorption in the two samples).
There is some evidence that sub-DLAs display larger metallicities than
DLAs \citep{DZ03, Pe05, Ku07}, and they have been suggested to be more
massive \citep{Kh07}. 

The structure of this paper is as follows. Sect. 2 covers the
observations, sample selection, and measurements. In Sect. 3
we present observed correlations in the data set. In Sect. 4 we
discuss the interpretation of these correlations, and we identify a
population of absorbers that may trace galactic winds. A summary is
presented in Sect. 5.

\section{Data acquisition and handling}
\subsection{Observations}
Our dataset was formed by combining the DLA/sub-DLA sample of
\citet{Le06} with the 
Hamburg-ESO DLA survey of \citet[][ 2007, in preparation]{Sm05}.
All the data were taken in the years 2000 to 2006 with the Very Large
Telescope/Ultraviolet-Visual Echelle Spectrograph (VLT/UVES), located
on the 8.2\,m VLT Unit 2 telescope (Kueyen) at Cerro Paranal, Chile. 
UVES is described in \citet{De00}.
The data reduction was performed as described in \citet{Le03},
using the interactive pipeline written by \citet{Ba00}, running on the
ESO data reduction system {\tt MIDAS}. The rebinned pixel size is
$\approx$2\,\kms\ and the data have a spectral resolution (FWHM) of
6.6\,\kms\ ($R$=45\,000).

\subsection{\cf\ sample selection}
We took the 81 DLAs and sub-DLAs in the raw sample with data 
covering \cf, and looked for \cf\ components in a range of
$\pm$1000\,\kms\ around the system redshift. 
Absorption line components were identified
as \cf\ if they were present in both $\lambda$1548 and $\lambda$1550 in
the correct (2:1) doublet ratio. 
In four cases the \cf\ lines were so contaminated by blends that
we rejected them from the sample. 
In three other cases, the \hi\ lines or neutral-phase metal lines were too
blended for a metallicity to be derived; these were also excluded.
No DLAs or sub-DLAs were found where \cf\ absorption is not present. 
The final sample contains \ndla\ DLAs
and \nsub\ sub-DLAs, which are listed in Table 1.
Five of the DLAs and two sub-DLA are at less than 5000\,\kms\ from the
QSO redshift, and so may be affected by radiation from the QSO
\citep[e.g.][]{El02}.
For completeness we retain these $z_{\rm abs}\approx z_{\rm qso}$
systems (also known as proximate systems) in the sample, but 
the corresponding data points are highlighted in all figures, and we
take note of any differences from the intervening population.

\subsection{Measurement}
\subsubsection{Apparent Column Density}
For each DLA and sub-DLA,
we fit a continuum to a region several thousand \kms\ in width
centered on \cf\ $\lambda$1548, using a polynomial fit (often linear)
to regions of the spectrum judged to be free from absorption. 
We defined the zero-point of the velocity scale using the 
redshift of the strongest component of neutral gas absorption.
We then determined $v_-$ and $v_+$, the velocities
where the \cf\ absorption recovers to the continuum on the blueward
and redward side of the line.
For each pixel between $v_-$ and $v_+$, the apparent
optical depth is defined as $\tau_a(v)={\rm ln}\,[F_c(v)/F(v)]$, where
$F(v)$ and $F_c(v)$ are the actual flux and the estimated continuum
flux as a function of velocity, respectively. The total apparent
optical depth is then found by integrating over the line,
i.e. $\tau_a=\int_{v-}^{v+}\tau_a(v){\rm d}v$.
The apparent column density in each absorber then follows by
$N_{\rm a}=[3.768\times10^{14}/(\lambda_0 f)]\tau_a $ \citep{SS91},
where $\lambda_0$ is in Angstroms, and $f$ is the oscillator strength
of the line. For the two \cf\ lines, we take $\lambda_0=1548.204,
1550.781$\,\AA\ and $f=0.1899, 0.09475$ from \citet{Mo03}; see also
\citet{PA04}.  
The apparent column density will equal the true column density 
so long as the lines are not heavily saturated, and that there is no
unresolved saturation in the line profiles.

\subsubsection{Total Line Width}
We require a precise measurement of the total \cf\ line width in each
system. Following \citet{PW97}, 
we define $\Delta v_{\rm \cf}$ as the velocity width that
contains the central 90\% of the integrated optical depth in the
line. By finding the two pixels where the cumulative integrated
optical depth is 5\% and 95\% of the total, and determining the
velocity difference between them, one obtains $\Delta v_{\rm \cf}$.
This can be done for each of the two lines in the \cf\ doublet, with
the same result expected in each case if the lines are unsaturated and
unblended. Note that $\Delta v_{\rm \cf}$ is integrated over all
components, and should not be
confused with the line widths of each individual component.

We also define an alternative measure of total line width as $v_+-v_-$, the
total velocity range over which \cf\ absorption is present, regardless
of saturation. 
$v_+-v_-$ is sensitive to weak but nonetheless interesting
satellite components that are not contained within $\Delta v_{\rm \cf}$.
These weak components are particularly relevant in the search for winds.
The drawback of using $v_+-v_-$ is that it has a larger error than 
$\Delta v_{\rm \cf}$, since $v_-$ and $v_+$ are 
selected by eye (we estimate
$\sigma_{v_+-v_-}=20$\,\kms) , and also that it is
sensitive to the signal-to-noise ratio (low optical depth absorption is
harder to detect in low S/N data). 
Our data is of high enough quality to ensure that the second effect should
not be a major concern: the noisiest spectrum in our sample has S/N=25, 
and the mean S/N is 51 (where the S/N is measured per
resolution element at the observed wavelength of \cf).

\subsubsection{Mean and Maximum \cf\ Velocity}
For each profile, we measure the average optical depth-weighted
velocity, denoted by $\bar{v}_{\rm \cf}$, and calculated by 
$\bar{v}_{\rm \cf}=\int^{v_+}_{v_-}v\tau_a(v)\mathrm{d}v/\int^{v_+}_{v_-}
\tau_a(v)\mathrm{d}v$. 
Since the velocity zero-point is defined by the
strongest absorption component in the neutral gas, 
$\bar{v}_{\rm \cf}$ is equivalent to the mean velocity offset
between the neutral and ionized gas. In the analysis we are only
concerned with the magnitude of the velocity offset, $|\bar{v}|$,
and not whether the gas is blueshifted or redshifted relative to the
neutral gas. We also make use of $v_{\rm max}$, the
maximum absolute velocity at which \cf\ absorption is observed (i.e.,
the terminal velocity), given by $v_{\rm max}={\rm max}(|v_-|, |v_+|)$. 

\subsubsection{Choice of Doublet Line}
The measurements of column density, total line width, and mean velocity were
conducted independently on $\lambda$1548 and $\lambda$1550.
To select which of the two \cf\ transitions to use for our final
measurement, we followed the following rules that assess the influence
of saturation. 
If the condition $F(v_0)/F_c(v_0)>0.1$ (corresponding to
$\tau_a(v_0)<2.3$) is true for $\lambda$1548, where $v_0$ denotes the
velocity where the absorption in strongest, we use $\lambda$1548 to
measure the \cf, otherwise we use $\lambda$1550.
If both lines are saturated (defined here as when 
$F(v)<\sigma_{F(v)}$ or $F(v)<0.03\,F_c(v)$ at any point
within the line profile), we proceed with a lower limit to the
column density and an upper limit to the line width 
using the results from $\lambda$1550.
However, if one of the two \cf\ lines is blended, we use the other
line for measurement, regardless of the level of saturation.
There are four cases 
where both \cf\ lines are partly blended, but we
still have useful information at other velocities within the line profiles.
This can occur when the two \cf\ lines, separated by
$\approx$500\,\kms, blend with each other.
In these cases, which are flagged in Table 1, we derived our best
estimate of log\,$N_{\rm \cf}$ using the sum of the column densities
measured over two separate unblended velocity ranges, and assuming
$\Delta v_{\rm \cf}=0.9(v_+-v_-)$ and $\bar{v}=(v_-+v_+)/2$,
where for these cases $v_-$ refers to the lower bound of absorption
of the lower velocity range, and $v_+$ refers to the upper bound of
absorption of the higher velocity range.

\subsubsection{Properties of the Neutral Phase}
In most of the systems in our sample, 
the metallicity of the neutral gas, the \hi\
column density, and the low-ion line width have
already been published in \citet{Le06}, so we take these measurements
directly from that paper. 
We also include measurements of [Zn/H] in two DLAs 
(at $z_{\rm abs}=2.34736$ toward \object{Q0438-0436}
and 2.18210 toward \object{Q2311-373}) from \citet{Ak05},
one DLA and one sub-DLA (at $z_{\rm abs}=1.85733$ and 1.87519,
respectively, toward \object{Q2314-409}) from \citet{EL01},
and two DLAs (at $z_{\rm abs}=2.40186$ toward \object{Q0027-186}
and 1.98888 toward \object{Q2318-111}) from \citet{No07}.
For the remaining cases where no neutral-phase measurements have been
published, we executed the measurements using exactly the same techniques
as in \citet{Le06}.
The metal line used to measure [Z/H] is \ion{Zn}{ii} if detected, otherwise
\ion{Si}{ii} or \ion{S}{ii}, and $N_{\rm \hi}$ is
derived from a fit to the damping wings of the \lya\ line.
Both zinc and silicon are found to be undepleted in DLAs \citep{PW02},
so these metallicities should not be significantly affected by dust.
We follow the standard practise of quoting metallicities on a
logarithmic scale relative to solar.
All measurements were adjusted to the solar reference levels
adopted by \citet{Mo03}.

\section{Results}
The measurements of \cf\ absorption in each of the \ntot\ systems are given
in Table 1, and the \cf\ profiles for each system (together with an
optically thin line showing the component structure in the neutral
phase) are shown in Fig. 1.
Our sample spans a redshift range from 1.75 to 3.61 with a median of 2.34.
The values of log\,$N_{\rm \hi}$ range from 19.70 to 21.80 with a median of
20.65, and the metallicity [Z/H] lies between $-$2.59 and $-$0.31 with
a median of $-$1.36 (i.e., approximately one twentieth of the solar value). 
The values of log\,$N_{\rm \cf}$ range from 13.02 to $>$15.41 (median value
14.15), with total line widths $\Delta v_{\rm \cf}$ between 35 and 1110\,\kms\
(median value 187\,\kms), and velocity offsets between 0 and 425\,\kms\
(median value 46\,\kms). 

Histograms of the total \cf\ line width (using both $\Delta v$ and
$v_+-v_-$) and \cf\ velocity offset are given in Fig. 2.
A significant difference between the distributions of $\Delta v$ and
$v_+-v_-$ can be seen, with the peak in the $\Delta v$
distribution at $\approx$80\,\kms, and the peak in the $v_+-v_-$
distribution occurring at $\approx$200\,\kms. This difference is due
to the presence of weak, outlying components which contribute to
$v_+-v_-$ but not to $\Delta v$. Both distributions show an extended tail 
reaching over 1000\,\kms. Fig. 2 also shows the distribution of mean
\cf\ velocities.

\setcounter{figure}{1}
\begin{figure}
\resizebox{\hsize}{!}{\includegraphics{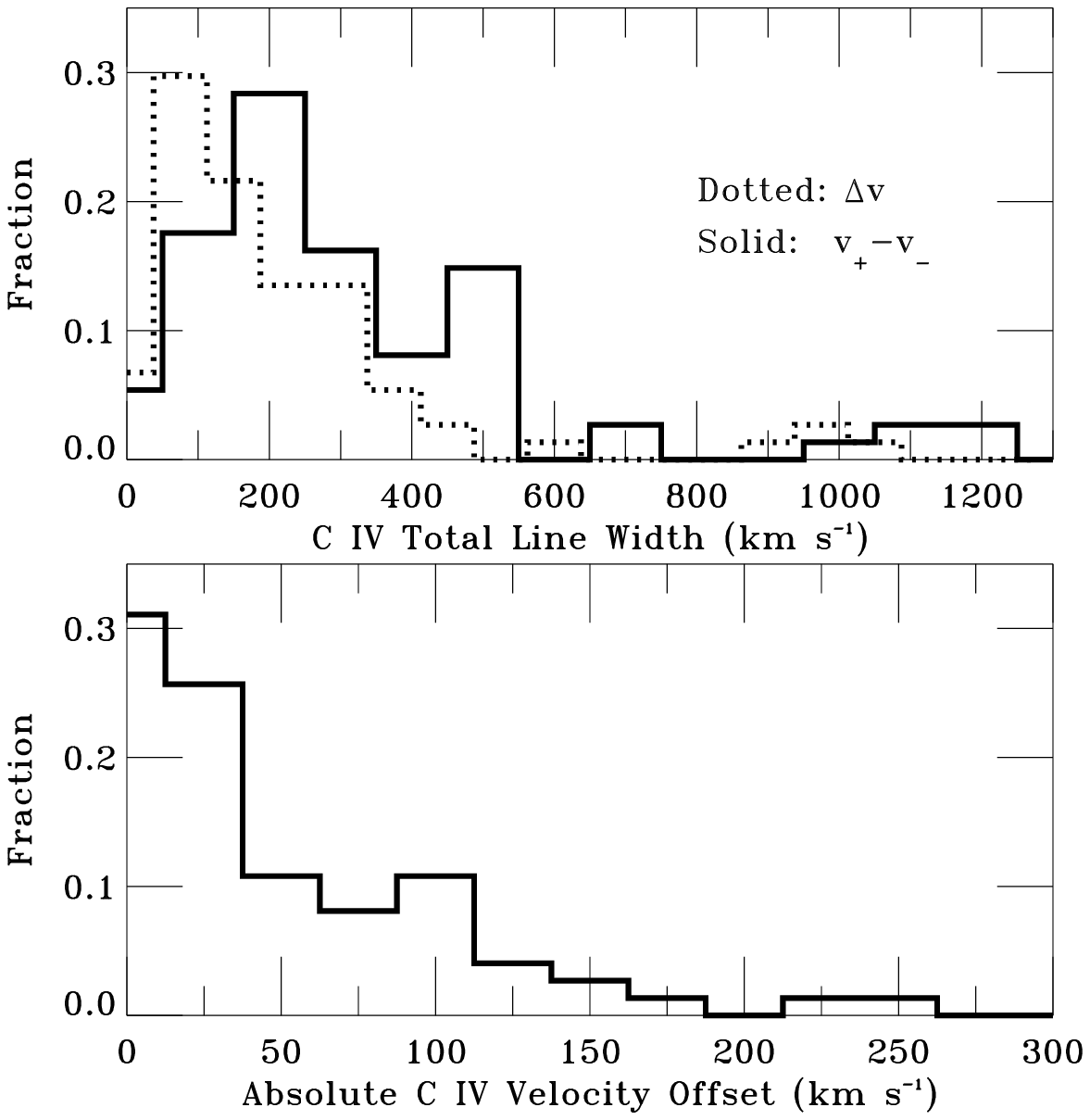}}
\caption{
  Normalized histograms of total \cf\ line width and absolute velocity
  offset among our DLA/sub-DLA sample. 
  The distribution of both $\Delta v$ (dotted) and $v_+-v_-$
  (solid) is shown in the top panel.
  The peak of the $v_+-v_-$ distribution is $\approx$100\,\kms\
  broader than the peak of the $\Delta v$ distribution, reflecting the
  presence of low optical depth satellite components.
  There is an extended tail of line widths reaching $>$1\,000\,\kms.
  We have treated the upper limits to $\Delta v$ as data
  points when forming the distribution.
  The distribution of the absolute \cf\ velocity offset (i.e., the
  mean \cf\ velocity relative to the neutral gas) is shown in the
  lower panel.
  }
\end{figure}

\subsection{DLA \cf\ vs IGM \cf}
\cf\ absorbers in the IGM falling at velocities within 1\,000\,\kms\
of the DLA/sub-DLA would not necessarily be physically connected to the
system. Such IGM \cf\ would contaminate our sample, in particular by
contributing to the high $\Delta v_{\rm \cf}$ tail shown in the top
panel of Fig. 2. IGM contamination could be occurring in systems 
where outlying, discrete components are seen at velocities
separated from the bulk of the absorption by hundreds of \kms\
(e.g, $z_{\rm abs}=2.293$ toward \object{Q0216+080}, 
$z_{\rm abs}=2.347$ toward \object{Q0438-436}, 
$z_{\rm abs}=1.825$ toward \object{Q1242+001}, and
$z_{\rm abs}=2.154$ toward \object{Q2359-022}).
However, we do not wish to exclude these absorbers from our sample,
since then we would be biased against finding high-velocity features,
such as winds. Our approach is thus to systematically include all \cf\
absorption in a fixed velocity interval around the system.

We can assess the level of IGM
contamination statistically by comparing the properties of the \cf\ in
our sample with those of \cf\ in the IGM. This is shown in Fig. 3,
where we compare our DLA/sub-DLA \cf\ column density distribution with
the IGM distribution at $z\approx$2--3 measured by 
\citet[][data taken from their Fig. 10]{Bo03}.
A similar IGM \cf\ distribution is presented by \citet{So05}.
We also include in Fig. 3 the distribution of \cf\ near LBGs, taken from
Table 3 in \citet{Ad05}; these measurements were 
made by finding LBGs lying at impact parameters of $<$1$h_{70}^{-1}$
co-moving Mpc from QSO sight lines, and then measuring the \cf\ column
densities in the QSO spectra at velocities within 200\,\kms\ of the
LBG redshift. Strong \cf\ absorption is also directly observed in LBG
spectra by \citet{Sh03}.

The DLA/sub-DLA \cf\ absorbers are clearly a different
population from the IGM \cf\ absorbers: the DLA/sub-DLA population shows a
mean column density that is higher by almost 1\,dex. The distribution of
\cf\ in DLAs and sub-DLAs resembles the distribution of galactic \cf\
as seen in LBGs, both with mean column densities near $10^{14}$\,\sqcm.
In consequence, the highest $N_{\rm \cf}$ systems in our sample are
\emph{least} likely to be of IGM origin. Since we report below that the
highest $N_{\rm \cf}$ systems tend to show the broadest \cf,  
we come to the conclusion that even the broadest \cf\ absorbers (that
were potentially the most suspect in terms of an association with
an individual galaxy), are likely to be galactic. 

\begin{figure}
\resizebox{\hsize}{!}{\includegraphics{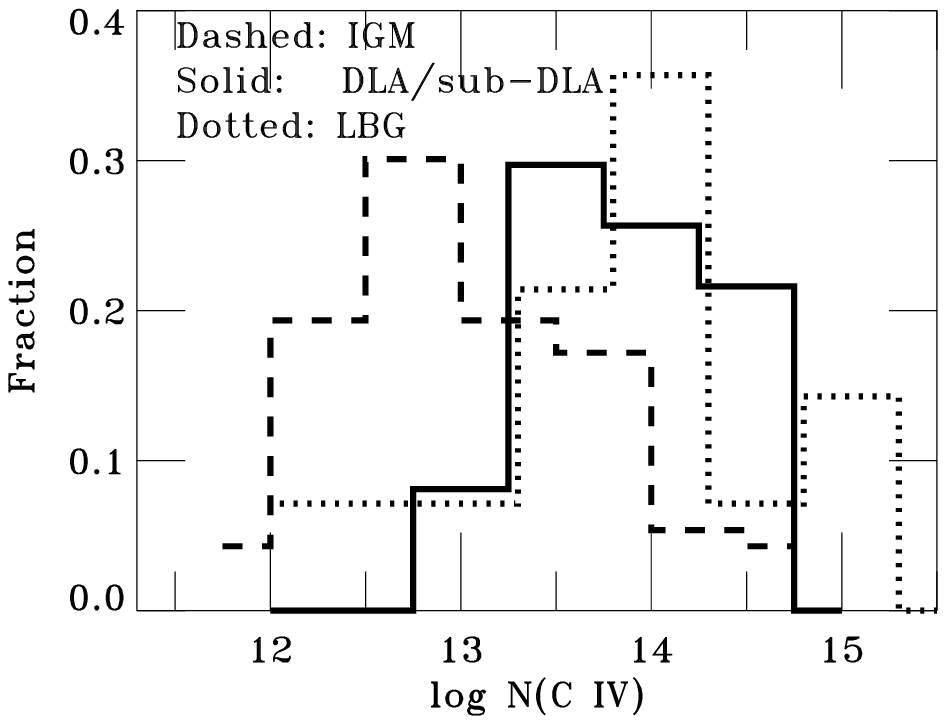}}
\caption{
 Comparison of the normalized \cf\ column density distributions: 
 (i) in DLAs and sub-DLAs (solid line, this work);
 (ii) in the IGM at $z\approx2-3$ \citep[dashed line;][]{Bo03};
 (iii) around LBGs at $z\approx2-3$ \citep[dotted line;][]{Ad05}.
 In each case, $N_{\rm \cf}$ is integrated over all components.
 Note how a typical DLA/sub-DLA shows (a) considerably stronger \cf\
 than a typical IGM \cf\ absorber, but (b) a \cf\ column similar
 to the mean seen in the LBG distribution. These two findings support
 our interpretation that the \cf\ in DLAs and sub-DLAs is galactic
 rather than intergalactic.
 We have included the saturated \cf\ absorbers as data points in the
 DLA/sub-DLA distribution, using the measured lower limits; this has
 the effect of artificially truncating the high $N_{\rm \cf}$ tail of
 the solid line.
}
\end{figure}

We now discuss correlations (or lack thereof) between the various
measured quantities in our dataset. For reference, a summary of all
correlations found and their statistical significance is given in Table 2.

\subsection{High-ion line width vs low-ion line width}
In Fig. 4 we directly compare the high-ion total line width with the
low-ion total line width for each DLA and sub-DLA in the sample. 
In almost all cases (69 of \ntot) 
the \cf\ lines cover a wider region of velocity space than the neutral
lines; this finding has been reported by \citet{Le98} and \citet{WP00a}.
We also find a considerable scatter ($\approx$1\,dex)
in $\Delta v_{\rm \cf}$ at low $\Delta v_{\rm Neut}$, but the scatter
decreases with increasing $\Delta v_{\rm Neut}$.

\begin{figure}
\resizebox{\hsize}{!}{\includegraphics{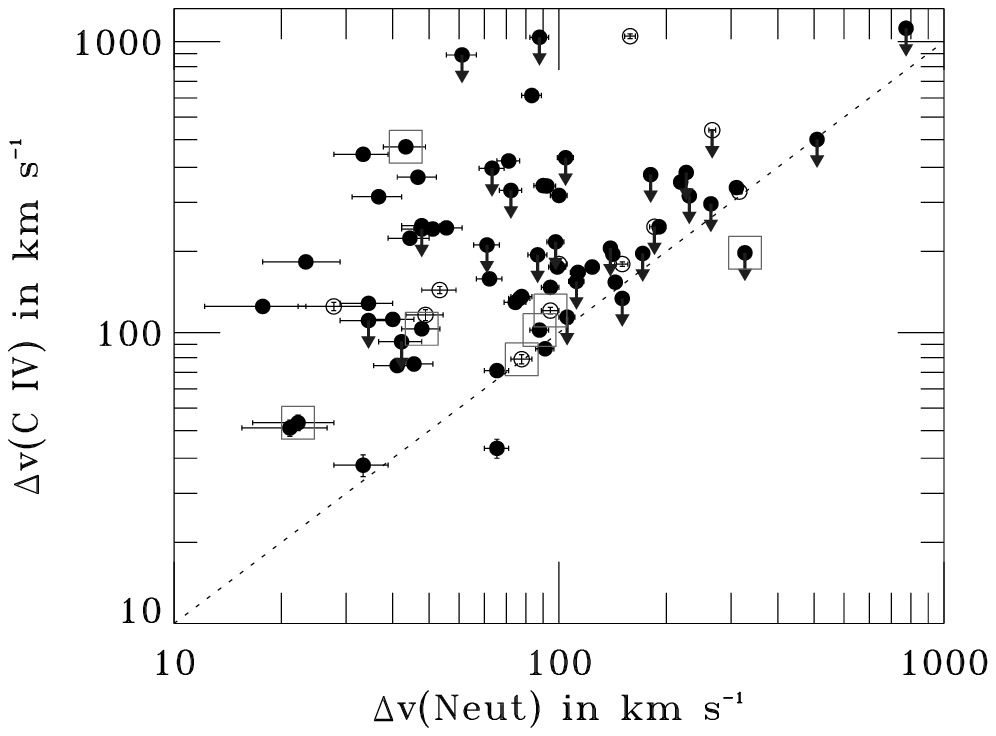}}
\caption{
  Comparison of high-ion and low-ion total line width for DLAs 
  (filled circles) and sub-DLAs (open circles). Absorbers at
  $<$5\,000\,\kms\ from the QSO redshift are highlighted in square symbols. 
  The dashed line shows where $\Delta v_{\rm \cf}=\Delta v_{\rm Neut}$.
  In 69 of \ntot\ cases the \cf\ absorption is broader than the neutral
  absorption. 
  There is a large scatter in $\Delta v_{\rm \cf}$
  at low $\Delta v_{\rm Neut}$, but
  the scatter decreases with increasing $\Delta v_{\rm Neut}$. 
  Saturated \cf\ absorbers are shown with upper limits to 
  $\Delta v_{\rm \cf}$.
  }
\end{figure}

\subsection{Internal correlations of \cf\ properties}
We find that the \cf\ column density, total line width, and
velocity offset are all correlated with one another. 
This is shown in Fig. 5, which
illustrates a $>$6.0$\sigma$ correlation between $N_{\rm \cf}$ and
$(v_+-v_-)_{\rm \cf}$, and a 4.3$\sigma$ correlation between
$\bar{v}_{\rm \cf}$ and $(v_+-v_-)_{\rm \cf}$. 
We investigated whether these two correlations were still found when
removing the proximate absorbers and the sub-DLAs from the sample, 
and found that they were, at $>6.0\sigma$ and 3.8$\sigma$
significance, respectively (see Table 2).
Finally, we considered the effect of the lower limits
on the $N_{\rm \cf}$ vs $(v_+-v_-)_{\rm \cf}$ correlation, by redoing
the analysis with the saturated points excluded. We still found a
correlation, but the slope (in log-log space) 
is lower by $\sim$0.6\,dex in this case (see Fig. 5). 
In summary, the DLAs and sub-DLAs with strong \cf\ absorption tend to show
broader and more offset \cf\ profiles. 

\begin{figure}
\resizebox{\hsize}{!}{\includegraphics{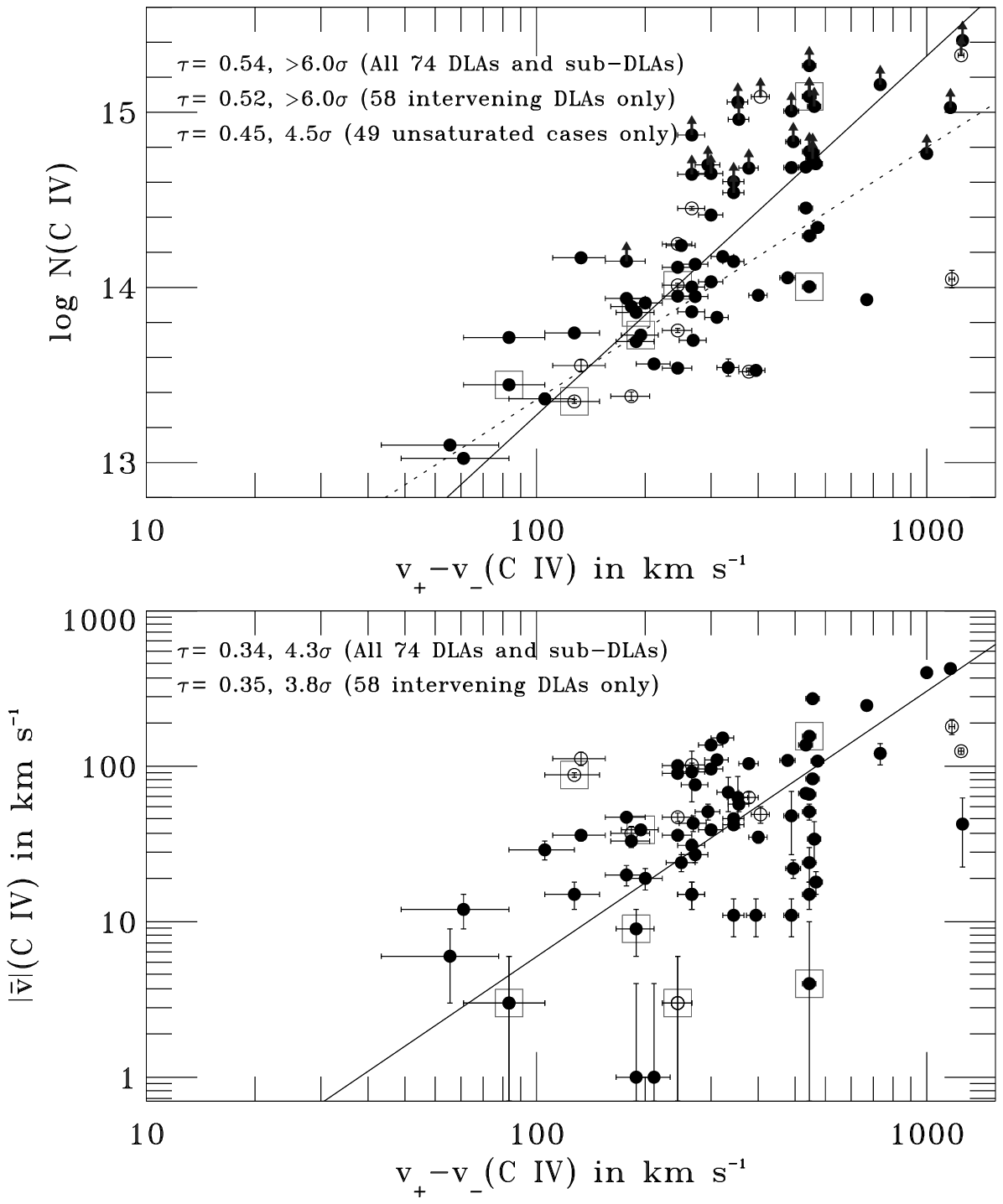}}
\caption{
 Correlations between the measured \cf\ properties
 for both DLAs (filled circles) and sub-DLAs (open circles). 
 Proximate absorbers are highlighted in square symbols.
 We use $v_+-v_-$ rather than $\Delta v$ to measure the line width,
 since it is defined even in the saturated cases.
 We annotate the Kendall rank correlation coefficient $\tau$
 and its significance on the panel, for various sub-samples. 
 Solid lines show linear least-square bisector fits for the case
 where all data points are treated equally (including limits).
 {\bf Top panel}: \cf\ column density vs \cf\ line width. 
 A correlation is found even when excluding the saturated \cf\
 absorbers (the lower limits on $N_{\rm \cf}$), although in this case
 the slope of the fit (shown as a dashed line) is shallower.
 {\bf Bottom panel}: comparison between total \cf\ line width and \cf\
 absolute velocity offset, also showing a significant correlation.
 These trends show that the stronger \cf\ absorbers tend to be broader
 and more offset from the neutral gas than the weaker absorbers.
  }
\end{figure}

\subsection{Metallicity vs \cf\ column density}
We plot $N_{\rm \cf}$ vs [Z/H] in Fig. 6. panel (a).
We emphasize that the metallicity is not derived from the \cf\ lines,
but is measured independently in the neutral phase of absorption,
using either the \ion{Zn}{ii}/\hi, \ion{Si}{ii}/\hi, or \ion{S}{ii}/\hi\ ratio.
A Kendall rank correlation test shows that the two
quantities are correlated at $>$6.0$\sigma$ significance (where the
limits were included in the analysis). 
Almost all the high-metallicity DLAs show saturated \cf\ lines, which
are shown with arrows to represent lower limits to $N_{\rm \cf}$. 
The correlation is still found (at 5.1$\sigma$) when only using the 
intervening DLAs, 
and is also detected (at 3.5$\sigma$) when only using the cases with
[Z/H] derived from \ion{Zn}{ii} (since these metallicities are more
robust against dust depletion effects). 
Finally the correlation is still found (but only at 2.5$\sigma$) when only
considering the unsaturated \cf\ data points (i.e., when ignoring the
lower limits). 
The detection of this correlation confirms the tentative result found
in Paper I based on a much smaller sample of twelve DLAs.

\begin{figure}
\resizebox{\hsize}{!}{\includegraphics{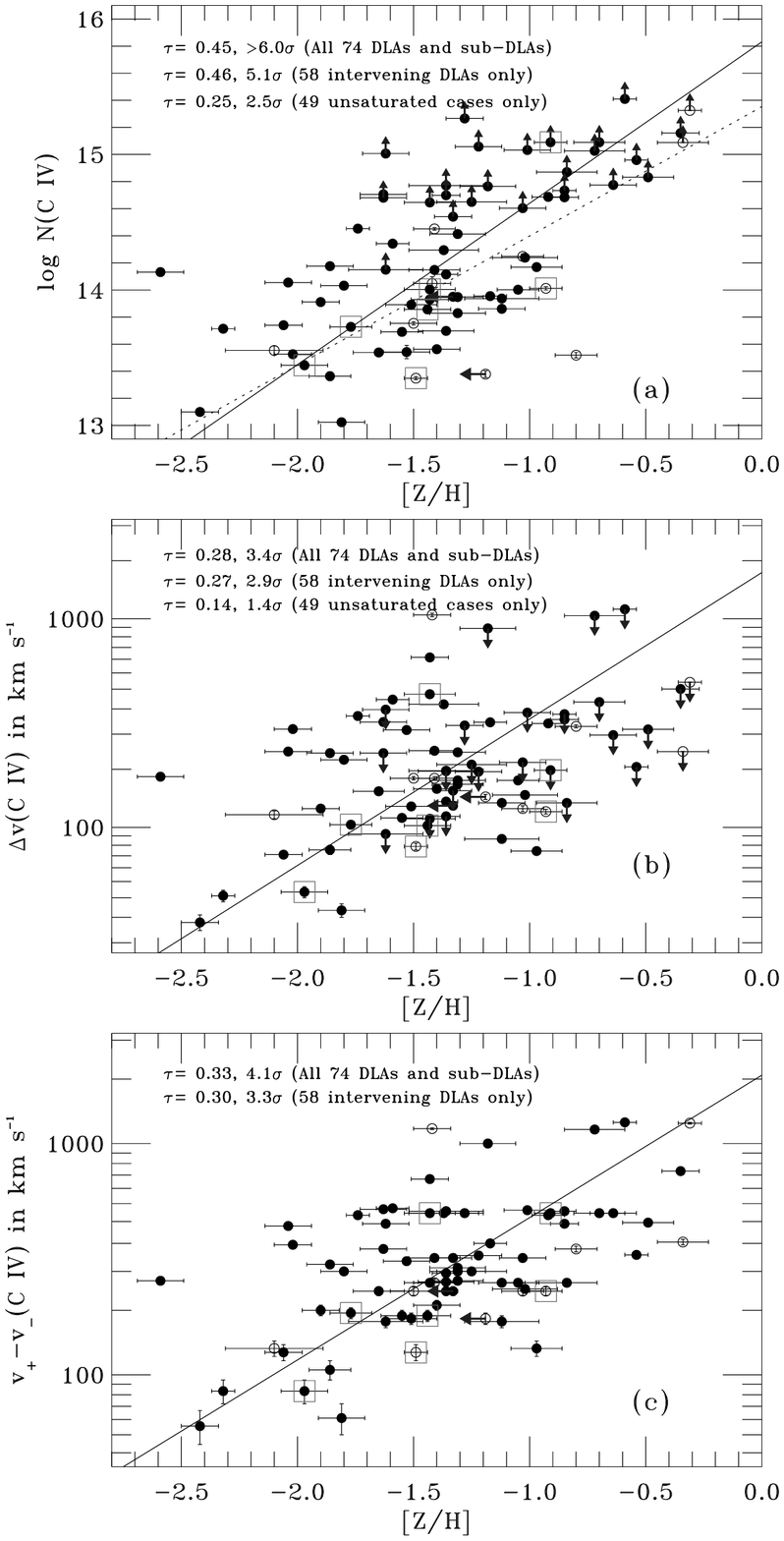}}
\caption{
  Dependence of \cf\ properties in DLAs (filled circles) and sub-DLAs
  (open circles) with neutral-phase metallicity. Proximate absorbers
  are highlighted in square symbols.
  In each panel, we annotate the Kendall rank correlation coefficient
  $\tau$ and its significance, and we show a linear least-square
  bisector fit (solid line), for the case where all data points are
  treated equally (including the limits).
  In panel (a), we find a significant correlation between 
  $N_{\rm \cf}$ and [Z/H]. This remains true (at 2.5$\sigma$) even in
  the case where the saturated points are excluded, though in this
  case the slope is slightly shallower (dashed line).
  In panel (b), we show a 3.4$\sigma$ 
  correlation between $\Delta v_{\rm \cf}$ and [Z/H], but this
  correlation is \emph{not} found when the saturated points are excluded
  (hence there is no dashed line in this panel). 
  However, if we instead use $(v_+-v_-)_{\rm \cf}$
  to measure the line width (bottom panel), 
  since this statistic is not affected by saturation,
  a significant (4.1$\sigma$) 
  metallicity-line width correlation does exist.
  The detected correlations show that high-metallicity systems tend to
  exhibit strong and broad \cf\ absorption. 
} 
\end{figure}
  
A linear least-squares bisector fit to the data gives the result:
\begin{eqnarray}
{\rm log}N_{\rm \cf}=(1.2\pm0.1){\rm [Z/H]}+(15.8\pm0.2) & [\rm{all\;points}]\\
{\rm log}N_{\rm \cf}=(1.0\pm0.1){\rm [Z/H]}+(15.4\pm0.2) & [\rm{no\;limits}]\nonumber,
\end{eqnarray} 
where $N_{\rm \cf}$ is expressed in \sqcm, the errors 
in the slope and y-intercept represent the 1$\sigma$ uncertainties,
and the first and second equations describe the cases where the limits
are included and excluded in the fit, respectively.

\subsection{Metallicity vs \cf\ total line width}
In Fig. 6 panel (b) we find evidence for a loose
correlation between metallicity and $\Delta v_{\rm \cf}$.
A Kendall $\tau$ test shows a 3.4$\sigma$ correlation when using
all \ntot\ DLAs and sub-DLAs. 
If we restrict the sample to the
\ndlaigm\ DLAs at $>$5\,000\,\kms\ from the QSO, to remove the effects
of sub-DLAs and proximity to the quasar, the significance of the
correlation decreases to 2.9$\sigma$. 
Working just with the DLAs and sub-DLAs with metallicities derived
from \ion{Zn}{ii}, the significance is 2.6$\sigma$. 

If we remove the upper limits on $\Delta v_{\rm \cf}$
(i.e., the saturated absorbers) from the sample, and redo the
correlation analysis, we find no significant detection of a
correlation between [Z/H] and $\Delta v_{\rm \cf}$ remains. 
However, in a sense this is not surprising, since the saturated
absorbers tend to show broad \cf\ lines (Sect. 3.4), 
so by removing them, we are biased against finding a trend with line
width. To further investigate whether saturation was 
playing a role in setting up this metallicity-line width relation, 
we looked for a correlation between [Z/H] and $(v_+-v_-) _{\rm \cf}$
As discussed in Sect. 2.3.2, $(v_+-v_-) _{\rm \cf}$ is defined even in the
saturated cases. The result was that we detected a positive
correlation at the 4.1$\sigma$ level. 
Since we believe this metallicity-\cf\ line width correlation to be
one of the most important results of this paper, we investigated
whether it was seen independently in the lower- and higher- redshift
halves of the sample, and found that it was
at $\approx3\sigma$ significance (Fig. 7), even though the mean
metallicity of the low-redshift sample is higher than the mean
metallicity of the high-redshift sample.
Together, these results imply that the \cf\
line width and metallicity are correlated in DLAs and sub-DLAs.

\begin{figure}
\resizebox{\hsize}{!}{\includegraphics{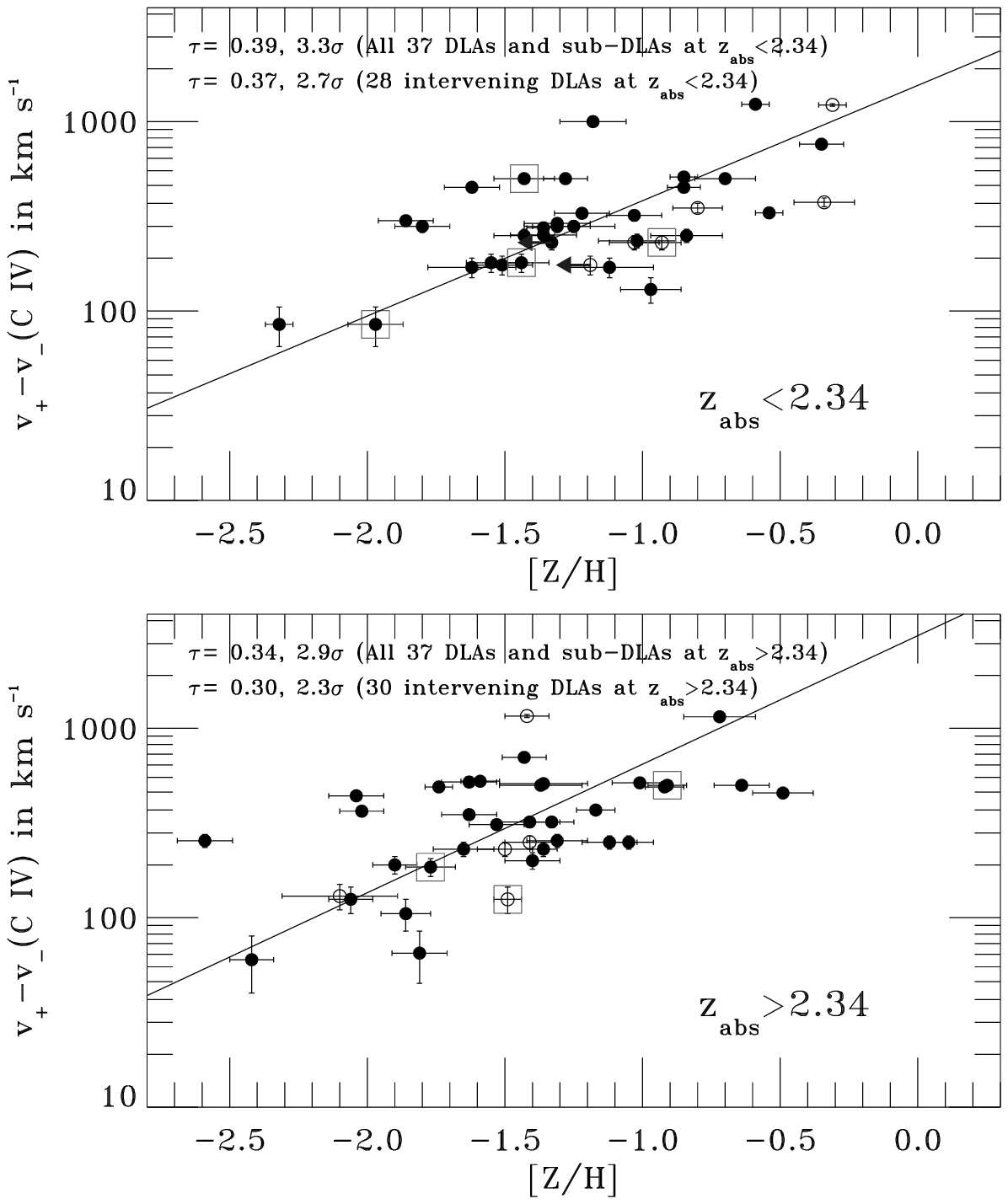}}
\caption{
  Illustration that the correlation between \cf\ total line width and
  metallicity exists independently in the lower and
  upper redshift halves of the sample, even though 
  there is a difference between the mean metallicity of the two
  sub-samples (the lower-$z$ sample shows systematically higher [Z/H]). 
  The symbols have their same meanings as in Fig. 6. 
  All DLAs and sub-DLAs in each redshift range were included in the
  correlation analysis and in the linear bisector fits, shown with
  solid lines.} 
\end{figure}

Using the sample of \ntot\ DLAs and sub-DLAs,
the best-fit linear least-squares bisector model is:
\begin{equation}
{\rm [Z/H]}=(1.4\pm0.1){\rm log}\Delta v_{\cf}-(4.6\pm0.5),
\end{equation} 
where again the errors are the 1$\sigma$ uncertainties, and where
$\Delta v_{\cf}$ is in \kms. The slope of this relation is similar to 
that found for the low-ion total line width/metallicity correlation by
\citet{Le06}, who report 
[Z/H]=(1.55$\pm$0.12)log\,$\Delta v_{\rm Neut}-(4.33\pm0.23)$. 
However, we observe a large dispersion in total \cf\ line width at a given
metallicity, far larger than the measurement errors,
and also larger than the dispersion seen in the low-ion line
width/metallicity correlation. 

There are four DLAs 
(at $z_{\rm abs}=2.418$ toward \object{Q0112+306},
$z_{\rm abs}= 2.076$ toward \object{Q2206-199}, 
$z_{\rm abs}= 2.153$ toward \object{Q2222-396}, and
$z_{\rm abs}=2.537$ toward \object{Q2344+125}),
and one sub-DLA (at $z_{\rm abs}= 3.170$ toward \object{Q1451+123})
that stand out on Fig. 6(b) and 6(c). 
due to their unusual properties. These absorbers show
narrow \cf\ lines ($\Delta v_{\rm \cf}<50$\,\kms\ in all cases), 
low velocity offsets ($|\bar{v}|_{\rm \cf}<10$\,\kms\ in four of five
cases), and low metallicities ([Z/H] between $-$2.42 and $-$1.81).
These cases play a significant
role in generating the correlations discussed here.

\subsection{Metallicity vs \cf\ velocity offset and terminal velocity}
Over our sample of \ntot\ DLAs and sub-DLAs, the mean value of
$|\bar{v}|_{\rm \cf}$ is 69\,\kms. 
$|\bar{v}|_{\rm \cf}$ is not correlated with the metallicity.
However, the absolute maximum \cf\ velocity
$v_{\rm max}$ is correlated with the metallicity at the 2.9$\sigma$
level (not shown in figures, but see Table 2). 
The significance of this correlation decreases to 2.6$\sigma$ when
just using the intervening DLA sample. 
Saturation has no effect on the maximum \cf\ velocity, so we included
the saturated \cf\ absorbers in this correlation analysis.
We note that $v_{\rm max}$ reaches over 200\,\kms\ in 42 cases, and
$>$500\,\kms\ in 8 cases. 

\subsection{\hi\ column density versus high ions}
We find that none of the \cf\ properties (column density, total line
width, and mean velocity) correlate with
$N_{\rm \hi}$, even though our sample covers two
orders of magnitude in $N_{\rm \hi}$. 
In Table 3 we compare the observed properties of \cf\ absorption in
DLAs (log\,$N_{\rm \hi}>20.3$) with
those in sub-DLAs (log\,$N_{\rm \hi}<20.3$). 
There is no significant
difference between the two populations in mean column density,
mean total line width, or mean velocity offset from the neutral gas.
However, if we assume that the ionization fraction \cf/C is the same
in all systems, then the DLAs tend to show larger \hw\ column
densities than the sub-DLAs (see Sect. 3.9). 

\subsection{Intervening vs proximate systems}
The differences between proximate DLAs and intervening DLAs have 
been studied in recent years \citep{El02, Ru06, HP07, Pr07b}.
Here we compare the properties of the \cf\ absorption in proximate
and intervening DLAs and sub-DLAs. These results are relevant to claims
that photoevaporation by the quasar reduces the \hi\ cross-section in
proximate DLAs \citep{HP07, Pr07b}. 
We find no evidence for a higher degree of ionization in the proximate
systems. Indeed, the mean \cf\ column densities and total line
widths in the proximate systems are slightly \emph{lower} than the
corresponding values in the intervening systems. However, our current
proximate sample with \cf\ only contains seven systems, so further
data are needed before strong conclusions can be drawn.

\subsection{Total ionized column density}
One of the key conclusions of Paper I was that, 
if the ionized and neutral phases of DLA have the same metallicity, then
the \hw\ column density in the \os\ phase typically amounts to
$>$40\% of the \hi\ column density in the neutral phase, and
that $N_{\rm \hw}$ in the \cf\ phase amounts to a further $>$20\% of
$N_{\rm \hi}$. These percentages are important since they determine the
total quantity of baryons and metals hidden in the ionized gas. With
the much larger sample in this paper, we are able to improve upon the
second of these estimates. $N_{\rm \hw}$ in the \cf\ phase is calculated by:

\begin{equation}
N_{\rm \hw}=\frac{N_{\rm \cf}}{f_{\rm \cf}{\rm C/H}}
=\frac{N_{\rm \cf}}{f_{\rm \cf}({\rm C/H})_\odot10^{\rm [Z/H]}}
\frac{Z_{\rm N}}{Z_{\rm I}}
\end{equation}

where $f_{\rm \cf}=N_{\rm \cf}/N_{\rm C}$ is the \cf\ ionization
fraction, and $Z_{\rm N}$ and $Z_{\rm I}$ are shorthands for the
absolute metallicities in the neutral and ionized gas.
We have assumed a solar elemental abundance pattern, so that
[C/H]=[Z/H], and we take the solar carbon abundance C/H=$10^{-3.61}$
from \citet{AP02}.
We assume that $Z_{\rm N}/Z_{\rm I}=1$, though values $<$1 are possible
in a scenario where metal-rich, ionized supernova ejecta has yet to
mix with the general interstellar medium (ISM), and values $>$1 are
possible in an accretion scenario. 
Finally, we assume that $f_{\rm \cf}=0.3$, since in Paper I we found this is
the maximum amount allowed in \emph{either} photoionization
\citep{Fe98} or collisional ionization \citep{GS07} models , and so it
gives the most conservative (lowest) value of $N_{\rm \hw}$.  
Lower values of $f_{\rm \cf}$ would increase the $N_{\rm \hw}$ estimates.

The resulting values of $N_{\rm \hw}$ are shown in Fig. 8.
Based on the \ntot\ systems in this sample, we find
that the mean and standard deviation of the warm ionized-to-neutral
ratio is 
$\langle {\rm log}(N_{\rm \hw}/N_{\rm \hi})\rangle=-1.0\pm0.6$, 
i.e. the \cf\ phase contains $>$10\% of the baryons of the neutral phase.
The sub-DLAs show a mean log\,$N_{\rm \hw}$ of 19.33, whereas the
DLAs show a mean log\,$N_{\rm \hw}$ of 19.77, a factor of $\approx$2.2
higher. 
This is because sub-DLAs show (on average) similar \cf\ columns as
DLAs, but higher metallicities. 
We note in the lower panel of Fig. 8 that there is no trend
for $N_{\rm \hw}$, which is $\propto N_{\rm \cf}$/(Z/H),
to depend on metallicity (although such a trend could be partly hidden by the
saturation of \cf\ in high metallicity systems).
Thus, the correlation reported in Sect 3.4 
between $N_{\rm \cf}$ and
[Z/H] appears to be a simple consequence of the metallicity alone, but
does not imply that there is more ionized gas (i.e. more \hw) in the
high-metallicity systems.
We also note that the scatter in $N_{\rm \hw}$ is substantial, of order 
$\approx$2\,dex in $N_{\rm \hw}$
at values of [Z/H] between $-$2.0 and $-$1.0.

\begin{figure}
\resizebox{\hsize}{!}{\includegraphics{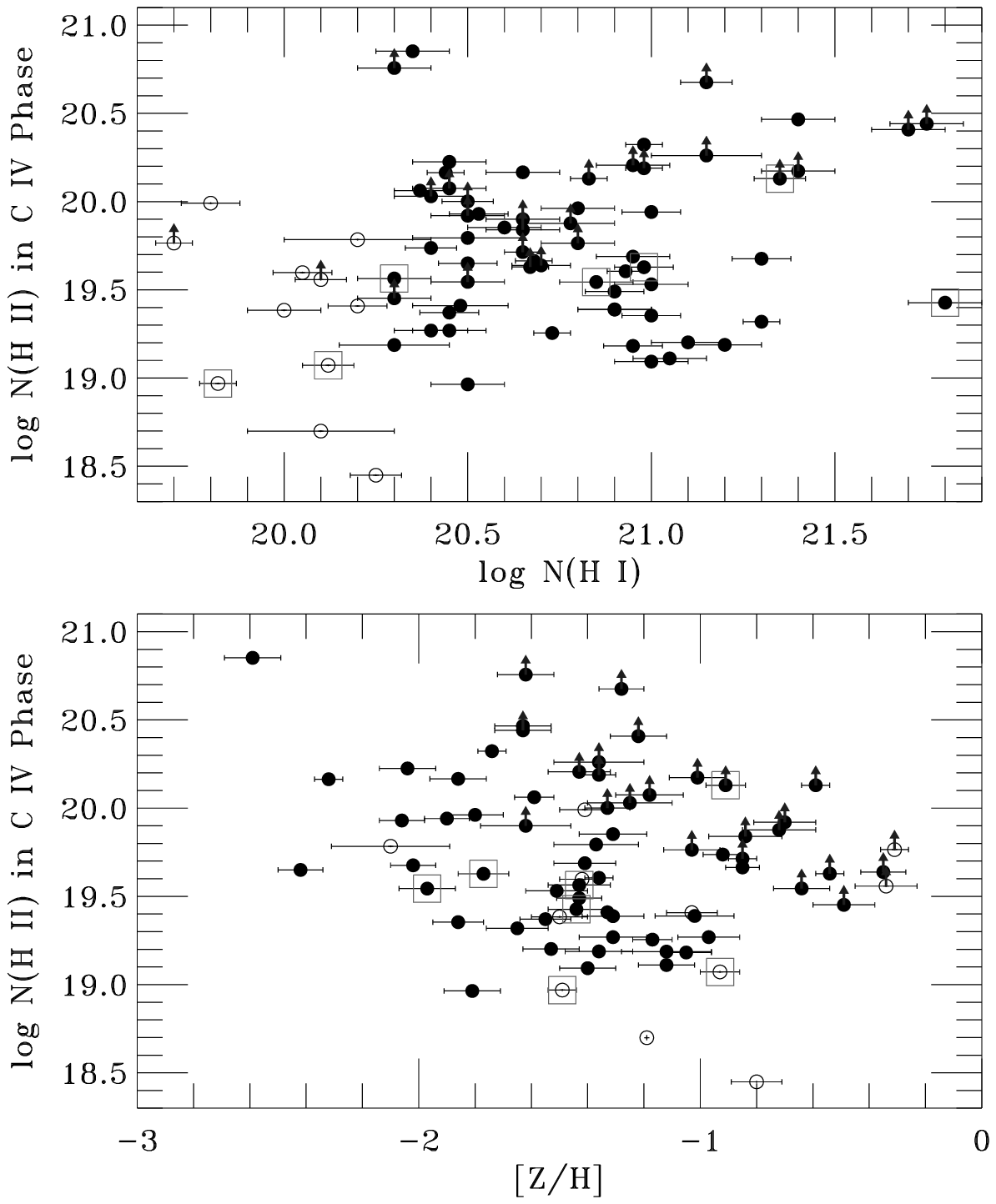}}
\caption{
 Comparison of \hw\ column density in the \cf-bearing
 gas integrated over all velocities
 with (top) $N_{\rm \hi}$ and (bottom) [Z/H] 
 for each system in our sample, assuming an ionization fraction
 $N_{\rm \cf}/N_{\rm C}=0.3$. 
 The average value of the ratio $N$(\hw)/$N$(\hi) over all \ntot\ systems is
 0.1, implying that the \cf\ phase of DLAs and sub-DLAs typically contains
 $>$10\% of the baryons and metals in the neutral phase. 
 The mean value of $N_{\rm \hw}$ is $\approx$2.5 times lower in
 sub-DLAs than in DLAs. 
}
\end{figure}

\subsection{\cf\ column density vs gas cooling rate}
The cooling rate $l_{\rm c}$ in diffuse interstellar gas is directly
proportional to the $N_{\rm \cw^*}/N_{\rm \hi}$ ratio\footnote{\cts\
  is an excited electronic state of singly ionized carbon.
  The transition at 158\,$\mu$m resulting from
  the decay from fine-structure state $^2P_{3/2}$ to $^2P_{1/2}$ 
  in the 2s$^2$2p term of C$^+$ is
  the principal coolant for diffuse neutral interstellar gas.},
according to 
$l_{\rm c}=N_{\rm \cw^*}h\nu_{\rm ul}A_{\rm ul}/N_{\rm
  \hi}$\,erg\,s$^{-1}$ per H atom, where $A_{\rm ul}$ is the Einstein
$A$ coefficient and $h\nu_{\rm ul}$ the energy of the 158\,$\mu$m line
\citep{Po79}. 
The cooling rate is of interest since it is equivalent to the
heating rate, because the cooling time is so shorter than the
dynamical time \citep{Wo03a, Wo03b}. In turn, the heating rate
will be related to the intensity of UV radiation and the dust-to-gas
ratio, and ultimately to the star formation rate per unit area.

We searched the literature for \cts\ measurements in our sample of
DLAs/sub-DLAs with \cf. We took 20 data points (11 measurements and 9 upper
limits) from \citet{Sr05}, 9 measurements from \citet{Wo03a}, and one
from \citet{He06}. In Fig. 9, we directly compare the cooling rate
with the \cf\ column density. 
Below log\,$l_{\rm c}=-26.8$, there is no trend evident in the data. 
However, we find that the seven points with the highest $N_{\rm \cf}$ 
are among the systems with the highest cooling rate.
Even though a formal correlation between the cooling rate and 
$N_{\rm \cf}$ is detected only at the 1.8$\sigma$ level (the \cts\
upper limits were excluded in this analysis), we note that the 
the median logarithmic \cf\ column density among the systems with 
log\,$l_{\rm c}<-26.8$ is 13.86, whereas
the median among the systems with 
log\,$l_{\rm c}>-26.8$ is 14.83. 
This finding is consistent with the results of
Wolfe (2007, in preparation), who finds 
evidence for bimodality in DLAs based on the cooling rate, 
in the form of significant differences between the metallicities
and velocity widths of those DLAs with cooling rates below and above a
critical value $l_c=10^{-27}$\,erg\,s$^{-1}$\,H$^{-1}$.

\begin{figure}
\resizebox{\hsize}{!}{\includegraphics{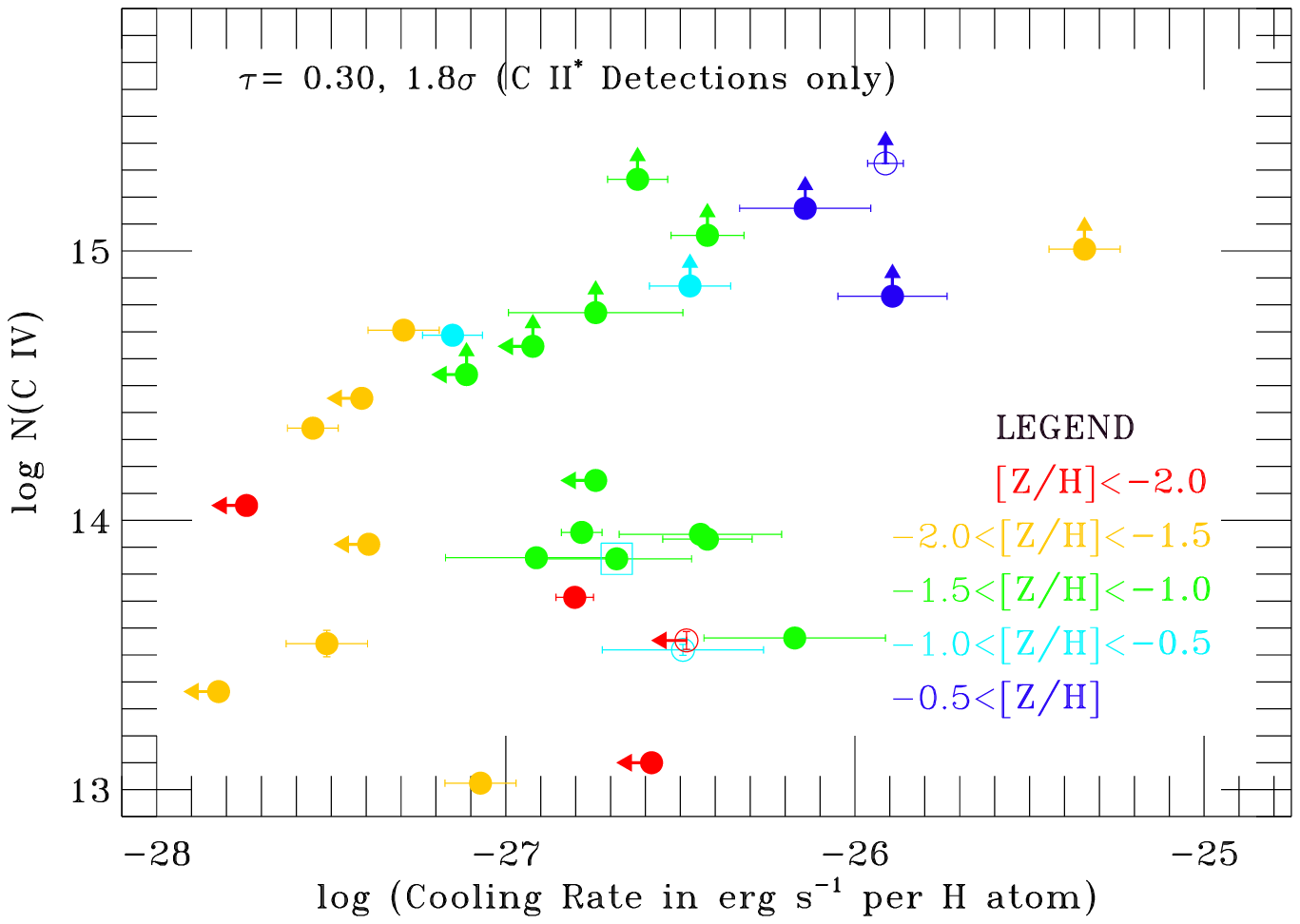}}
\caption{Dependence of the \cf\ column density on the 
  cooling rate, derived from the $N_{\rm \cw^*}/N_{\rm \hi}$ ratio,
  for each DLA (filled circles) and 
  sub-DLA (open circles) where data on \cf\ and \cts\ exist. 
  Color-coding is used to denote the metallicity of the gas, as
  indicated in the legend.
  The seven data points with the highest $N_{\rm \cf}$ all show 
  above-average cooling rates. Among these seven,
  six show high metallicities. 
}
\end{figure}

\section{Discussion}

\subsection{Narrow and broad components}
There are two physical processes that can provide the 47.9\,eV
required to ionize C$^{+2}$ to C$^{+3}$, and so create the gas seen in
\cf: photoionization and collisional ionization.
In Paper I it was shown [see Fig 3(b) in that paper] 
that the line widths of at least one
fifth of the \cf\ components observed
in DLAs are narrow ($b\la10$\,\kms), implying that in these
components, the gas is cool ($T<7\times10^4$\,K), which likely implies
collisional ionization is unimportant, and that photoionization is  
the ionization mechanism \footnote{Cool but highly ionized clouds
  could however be produced by a non-equilibrium collisional
  ionization scenario, in which initially hot gas has cooled faster
  than it has recombined, leading to ``frozen-in'' ionization \citep{Ka73}.}.
These narrow components are not seen in \os. 
The detection of cool \cf\ components rules out the idea
that {\emph all} the \cf\ in DLAs arises in a hot halo
\citep[see][]{Mo96, Ma03}.
The source of the extreme-ultraviolet (EUV) radiation at 259\,\AA\
that photoionizes C$^{+2}$ to C$^{+3}$ and gives rise to the narrow
\cf\ components could be external (the extragalactic background)
or internal (O- and B- type stars in the DLA host galaxies).
Discussions of the relative importance of internal and external
radiation in DLAs are given by \citet{HS99}, \citet{ME05}, and \citet{Sy06}.
Note that in the Milky Way, \citet{BH86} found that planetary
nebulae are the dominant source of photons in the range 45--54\,eV,
but DLA galaxies at $z>2$ are likely too young for planetary nebulae to
have formed.

We propose that the broad \cf\ components arise in the hotter phase of DLA
plasma that is detected in \os\ absorption (Paper I), i.e the hot
ionized medium. This phase will arise following either heat input from
supernova in the DLA host galaxy, or by the shock-heating of
infalling gas at the virial radius. In the first case, the hot ionized
medium may exist in the form of a wind \citep{OD06, Fa07, KR07},
though Galactic fountain scenarios are also possible \citep{SF76, Br80, HB90}.
The observation that up to 80\% of the \cf\ components are
broad is consistent with the origin of the \cf\ in a wind,
since in the models of \citet{OD06}, much of the 
\cf\ in galactic winds is collisionally ionized.

We note that Type II supernovae will heat interstellar gas to temperatures
$>$10$^6$\,K, too high for the formation of \os\ and \cf\ lines, and
left to itself, gas at a density of 10$^{-3}$\,cm$^{-3}$ and
one-hundredth of the solar metallicity will not cool in a Hubble time.
However, if the hot plasma interacts with cool or warm entrained clouds,
conductive interfaces \citep{Bo90}, turbulent mixing layers
\citep{Sl93, Es06}, or shock fronts \citep{DS96}
can form between the hot and cool phases, in which
the temperatures are favorable for the formation of \os\ and \cf\
lines. These mechanisms, which have been invoked to explain high-ion
observations
in the extended halo of the Milky Way \citep{Sa03, Zs03, IS04}, 
in high-velocity clouds \citep{Se03, Fo04, Fo05, Co05}, and in the Large
Magellanic Cloud \citep{Le07}, can explain the broader \cf\ and \os\
components seen in the DLAs and sub-DLAs.
Note that the interpretation of broad high-ion components in
DLAs and sub-DLAs as hot and collisionally ionized is different from the
interpretation of the \os\ components in the IGM at $z\ga2$, 
which (generally) appear to be photoionized \citep{Ca02, Be02, Lv03,
Be05, Re06, Lo07}, though see \citet{Re01} and \citet{Si02, Si06}. 

\subsection{Ionized gas and star formation}
\citet{Le06} have presented a [Z/H]-$\Delta v_{\rm Neut}$
correlation in DLAs and sub-DLAs, and have interpreted it as implying
an underlying mass-metallicity relation \citep[see also][]{WP98, Le98,
Mu07, Pr07a}.
In this interpretation, $\Delta v_{\rm Neut}$ traces motions due to gravity.
Since we find that $\Delta v_{\rm \cf}$ is larger than 
$\Delta v_{\rm Neut}$ in almost all cases, an additional energy source
is required to heat and accelerate the \cf\ clouds.
We suggest that star formation and subsequent supernovae could provide
this source. Star formation in DLA and sub-DLA galaxies will lead to:\\
(i) metals generated by stellar nucleosynthesis;\\ 
(ii) EUV flux from OB stars that can photoionize C$^{+2}$ to C$^{+3}$ in
interstellar gas, giving rise to the narrow \cf\ components;\\
(iii) supernovae-heated million-degree plasma, which can interact with
entrained clouds of cooler gas to produce gas at $T\sim10^5$\,K, where
\os\ and \cf\ components are formed through electron collisions;\\ 
(iv) mechanical energy injection from supernovae and stellar winds
imparting large total velocity widths to the high ions.\\
Because star formation leads to metals and to feedback
(i.e. mechanical energy injection into the ISM), this scenario would
naturally explain the correlation between [Z/H] and $\Delta v_{\rm \cf}$.
However, we cannot rule out an alternative scenario in which the
plasma phases in DLAs and sub-DLAs are formed following the accretion
of infalling, intergalactic gas, rather than by star formation
\citep[e.g.][]{WP00b}. The inflow model can also qualitatively explain the
metallicity-\cf\ line width correlation: the more massive halos 
(which through the mass-metallicity relationship tend to show higher
metallicities) would induce higher accretion rates because of their
deeper potential wells, and so could create and disperse the \cf\ over
large velocity ranges. 

\subsection{Evidence for galactic outflows and winds}
Our dataset shows that \cf\ components in DLAs and sub-DLAs exist over
a broad velocity range, with a median $\Delta v_{\rm \cf}$ of
187\,\kms, which is approximately twice as broad as the typical
velocity spread seen in the neutral gas.
The terminal \cf\ velocities reach $>$200\,\kms\ in 42/\ntot\ systems,
and $>$500\,\kms\ in 8/\ntot\ systems, and are correlated with the
metallicity. Together, this evidence implies that the high ions in
DLAs and sub-DLAs trace highly disturbed kinematic environments.
In this section we identify a population of high-velocity \cf\
components with intriguing ionization properties, and
we address whether these components could be created by galactic outflows.

In order to evaluate whether any of the observed \cf\ components
represent winds, we need to determine the escape velocity in each
system. We calculate this using
$v_{\rm circ}=\Delta v_{\rm Neut}/0.6$, an empirical relation found from
analysis of artificial spectra in the simulations of \citet{Ha98} and
\citet{Ma01}. There is a factor of two dispersion around this
relation, due to variations in the viewing angle.
We then take $v_{\rm  esc}=\sqrt2v_{\rm circ}$ (appropriate for a
spherical halo), so that $v_{\rm esc}\approx2.4\Delta v_{\rm Neut}$.
The escape speed we have assumed can be taken as an upper limit,
because $v_{\rm esc}$ is calculated in the disk and will decrease with
radius, and we may be observing \cf\ at high radii.
We have not accounted for drag forces arising due to entrainment
between the winds and the galaxy's ISM \citep{FD07}.

When we search for \cf\ absorption at $v<-v_{\rm esc}$ and $v>v_{\rm esc}$, 
we find it in 25 of \ndla\ DLAs and 4 of \nsub\ sub-DLAs, i.e. 
\cf\ absorption unbound from the central potential well
exists in $\approx$40\% of cases. 
These absorbers, which we refer to as wind candidates, are colored with dark
shading in Fig. 1, for easy identification. 
A key property of the wind candidate absorbers is the low column
density of the accompanying neutral gas absorption.
In this respect, the wind candidates are analogous to the highly
ionized high-velocity clouds seen in the vicinity of the Milky Way
\citep{Se95, Se99, Se03, Co04, Co05, Fo05, Fo06, Ga05}.
They also resemble the $z\approx6$ \cf\ absorbers reported by
\citet{RW06}. 
In a handful of cases, wind candidates are seen at both redshifted and
blueshifted velocities in the same system.
In Fig. 10 (top panel) we
plot the absolute wind \cf\ columns as a function of metallicity.
We find similar \cf\ wind column densities in systems that span the
2.5\,dex range of metallicity in our sample, even in the
highest metallicity systems.

Our results are surprising when viewed in the light of simulations of
galactic outflows, which show that dwarf galaxies are more
important than massive galaxies for the metal pollution of the
intergalactic medium 
\citep{MF99, Fe00, Na04b, Sc06b, Ti06, Ko07}, 
since they are incapable of gravitationally confining the
metals released by supernovae. The increase in wind escape fraction
with decreasing galactic mass may contribute to the
origin of the mass-metallicity relationship observed in DLAs and other
high-redshift galaxies \citep{Na04a, Tr04, Mo04, Sa05,
  Er06}\footnote{Low star formation efficiency at low galactic masses
  may also contribute to the mass-metallicity relationship
  \citep{Br07, FD07}.}.
The surprising result here is that, given the mass-metallicity
relation, the higher metallicity (higher mass) galaxies should
decelerate their supernova-driven outflows, so that \cf\ outflows should
not be seen in the high-metallicity systems, but yet we observe the
high-velocity components even in systems with [Z/H]$>-1.0$. 
Furthermore, we find that the maximum outflow
velocity $v_{\rm \max}$ is correlated to the metallicity.

For any individual \cf\ component, it is difficult 
to determine conclusively whether one is seeing a galactic outflow
from the DLA galaxy \citep[see][]{Fa07}. 
Two other plausible origins are inflow toward the DLA galaxy
\citep{WP00b}, and the ISM of a separate nearby galaxy. 
A contribution of \cf\ from these processes
could help to explain both the scatter seen in each
panel in Figure 6, and the cases with large values for $\Delta v_{\rm \cf}$.
One weakness of the inflow model is that accretion would accelerate
the gas up to but not beyond the escape velocity, 
but yet we see gas moving \emph{above} the escape velocity.
Thus inflow cannot explain the highest-velocity \cf\ components.
The nearby galaxy model has the problem of not readily explaining why the 
ionized-to-neutral gas ratio is so high in the high-velocity
\cf\ components. In other words, if nearby galaxies are
responsible for the high-velocity \cf\ components, where is their
neutral ISM? 
The outflow explanation, on the other hand, naturally explains the
velocities and the ionization properties of the high-velocity \cf\ absorbers.
The outflow model also explains the presence of
metals in the ionized gas (they came from the DLA host galaxy), so it
does not need to resort to pre-enrichment.
A more serious problem is whether we can associate a single DLA with a
single galaxy. This assumption may be false since the clustering
of galaxies near DLAs has been observed both at low \citep{CL03} and
high \citep{Co06a, Co06b, BL03, BL04, El07} redshift.
Our wind interpretation implicitly assumes that each high-velocity \cf\
component arose from a galaxy located in velocity at the point where
the neutral line absorption is strongest. 

\begin{figure}
\resizebox{\hsize}{!}{\includegraphics{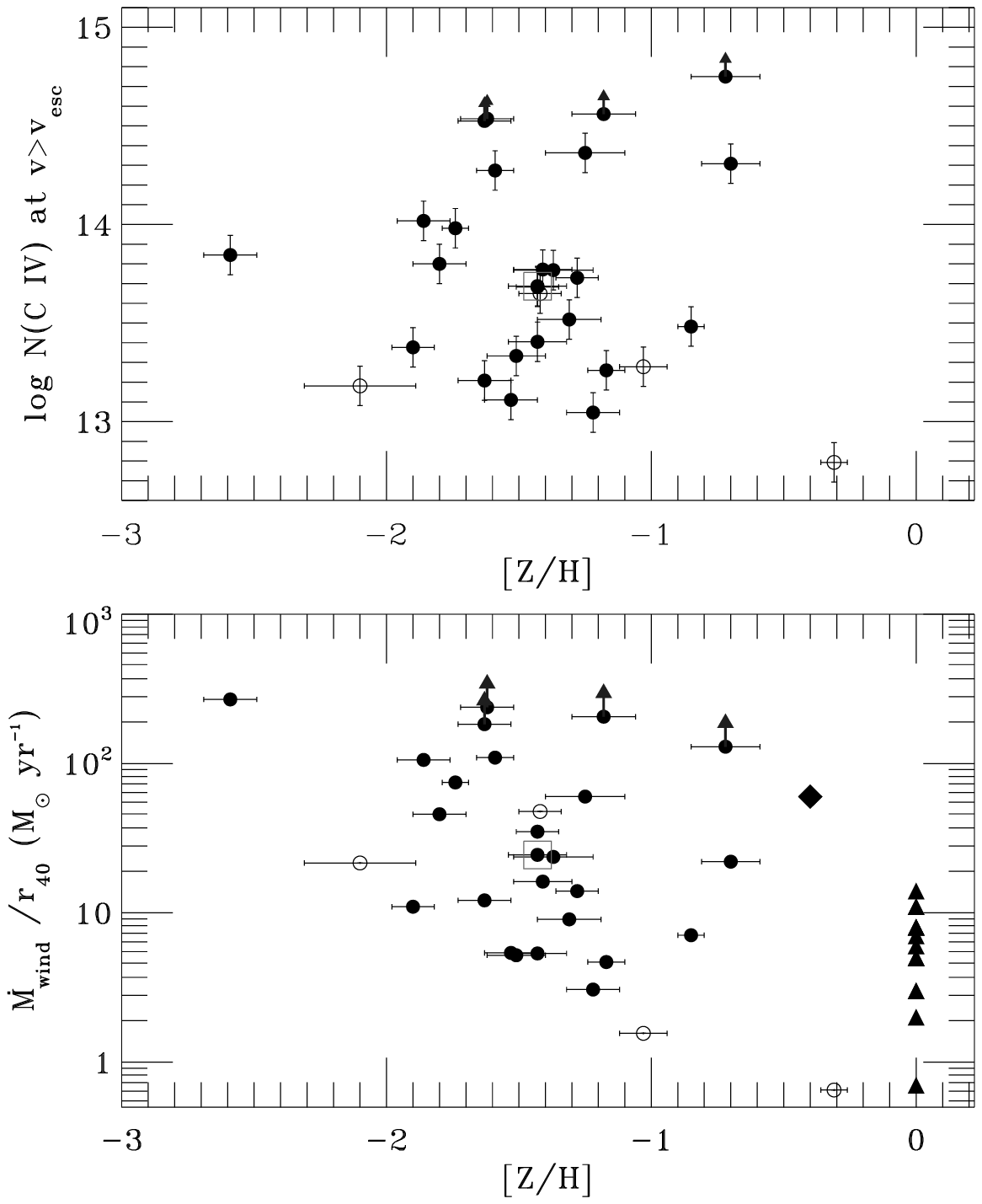}}
\caption{
  {\bf Top panel}: \cf\ column density moving above the escape speed 
  (i.e. wind candidate $N_{\rm \cf}$) as a function of metallicity, for the 25
  DLAs and 4 sub-DLAs with \cf\ absorption at 
  $|v|>|v_{\rm esc}|$, where $v_{\rm esc}=2.4\Delta v_{\rm Neut}$.
  {\bf Bottom panel}: Wind mass outflow rate $\dot{M}_{\rm wind}$ 
  divided by characteristic \cf\ radius $r_{40}$ (in units of 40\,kpc)
  vs metallicity. 
  The median mass outflow rate among these cases is 
  22$\langle r_{40} \rangle$\,\mfu. 
  The large diamond shows the mass outflow rate determined in the
  $z=2.7$ Lyman Break Galaxy cB58 by \citet{Pe00}, assuming $r$=1\,kpc.
  The filled triangles are the mass flow rates determined for the
  cool winds in the \citet{Ma06} sample of low-redshift ultraluminous
  starburst galaxies and plotted at zero metallicity for convenience.
  Only the DLA and sub-DLA points have been divided by $r_{40}$. 
}
\end{figure}

\subsubsection{Wind properties}
We present in this section a calculation of
order-of-magnitude estimates for the mass $M$, kinetic energy
$E_{\rm k}$, mass flow rate $\dot{M}$, and flux of kinetic energy
$\dot{E}_{\rm k}$ in the \cf\ wind candidate absorbers.
These calculations are analogous to those used to measure the
energetics of winds in starburst galaxies \citep{He00, Ma05, Ma06} and
LBGs \citep{Pe00}.

We assume that the outflow exists in an expanding hemispherical shell
moving at velocity $v$ relative to the neutral gas in the DLA galaxy
(we do not assume a full spherical outflow, since we do not generally
see two components corresponding to the near and far side of the galaxy). 
We then calculate $M$, $E_{\rm k}$, $\dot{M}$, and $\dot{E}_{\rm k}$
using the following equations:
\begin{eqnarray}
M & = & 2\pi r^2\mu m_{\rm H}N_{\rm \hw}\\ 
\dot{M} & = & 4\pi\mu m_{\rm H}r|\bar{v}|N_{\rm \hw}\\ 
E_{\rm k} & = & M\bar{v}^2/2\\
\dot{E_{\rm k}} & = & \dot{M}\bar{v}^2/2
\end{eqnarray} 

Here $\mu=1.3$ is the mean molecular weight, $m_{\rm H}$ is the mass of
a hydrogen atom, and $N_{\rm \hw}$ is calculated using Eqn. 3.
It can be seen that these four quantities are proportional to the zeroth,
first, second, and third moments of the \cf\ optical depth profile.
The wind mass flow rates are shown in the lower panel of Fig. 10.
They were calculated by taking the average
of the escape velocity and the terminal velocity for $\bar{v}$, and
deriving $N_{\rm \hw}$ with Eqn. 3 only for that portion of the \cf\
absorption at $|v|>v_{\rm esc}$. 
We only have information on the line-of-sight
velocity (not the full 3D velocity), but since the
line-of-sight passes through a DLA or sub-DLA, we assume we are looking
through the full depth of the galactic halo, so that the line-of-sight
velocity we measure approximates the radial velocity of the outflowing gas. 

The only quantity which is not directly measured is the characteristic
radius $r$. If we take a reference value of 
$r=40$\,kpc, as determined for \cf\ around LBGs by \citet{Ad05}, we find that 
the 29 \cf\ wind candidates typically contain (median values)
a total mass of $\sim2\times10^9$\,M$_\odot$, 
a kinetic energy of $\sim1\times10^{57}$\,erg, 
a mass flow rate of $\sim22$\,\mfu, 
and a kinetic energy injection rate of $\sim4\times10^{41}$\,erg\,s$^{-1}$. 
To put these wind mass flow rates in perspective,
\citet{Pe00, Pe02} report a mass flow rate of $\approx$60\,\mfu\ for the
outflow from the $z=2.7$ LBG MS~1512-cB58, and 
\citet{Ma06} calculate the mass loss rates in the cool
winds of a sample of low-redshift ultraluminous starburst galaxies as
between 1 and $>$14\,\mfu.

We calculate that the thermal energy in the \cf\ wind absorbers is
only a few percent of their kinetic energy (assuming a temperature of
$10^5$\,K).
Assuming the clouds fill a region whose depth is similar to its width,
a 40\,kpc depth and $N_{\rm \hw}=10^{20}$\,\sqcm\ correspond to a density of 
$n_{\rm \hw}\sim10^{-3}$\,cm$^{-3}$, assuming a filling factor of
unity. In turn, such a medium would exhibit a thermal pressure
$P/k\sim10^2$\,cm$^{-3}$\,K. If the \cf\
arose in interface layers between entrained cool/warm clouds and a
surrounding hot gas, then the line-of-sight filling factor would be
much lower and the \cf\ could exist in higher-density,
higher-pressure, localized regions.

Among the 25 DLAs and 4 sub-DLAs showing wind candidate components, 
the median rate at which each galaxy delivers metals to the IGM is
$\sim1\times10^{-2}$\,\mfu. 
This estimate is robust to changes in the metallicity of the ionized
gas with respect to the neutral gas.
Assuming that the outflows are driven by the supernovae that follow
star formation, we can calculate the star formation rate 
necessary to power the observed \cf\ outflows.
We use the relation from \citet{Ef00} between kinetic energy injection
rate and star formation rate, 
$\dot{E}_{\rm SN}=2.5\times10^{41}\dot{M}_\star$\,erg s$^{-1}$, 
with $\dot{M}_\star$ in \mfu~\footnote{This relation assumes each supernova
releases $10^{51}$\,erg of kinetic energy and that one supernova
follows from every 125\,$M_\odot$ of star formation.}.
Equating $\dot{E}_{\rm SN}$ to the observed mean flux of kinetic energy, we
find a median required SFR per DLA of $\sim$2\,\mfu, 
corresponding (with our assumed $r=40$\,kpc)
to a required SFR per unit area ($\dot{\psi_*}$) of 
$\sim3\times10^{-4}$\,\mfu\,kpc$^{-2}$. 
This is consistent with (though close to) the limit on the SFR per
unit area derived from low surface brightness features (i.e. DLA
analogs) in the Hubble Ultra-Deep Field by 
\citet{WC06}, who report $\dot{\psi_*}\la10^{-3.6}$\mfu\,kpc$^{-2}$. 
We thus conclude that, at least to an order-of-magnitude, there is
sufficient energy released following star formation to drive the
observed high-velocity \cf\ components in DLAs, supporting the notion
that they trace galactic winds.

\section{Summary}
We analyzed the \cf\ absorption in \ndla\ DLAs and \nsub\ sub-DLAs
in the range $1.75\!<\!z_{\rm abs}\!<\!3.61$ observed with VLT/UVES.
We measured the properties of the \cf\ absorption in each system and
investigated how they depend on the properties of the neutral gas.
In most systems the metallicity, \hi\ column density, and line width
in the neutral gas had already been measured.
We executed the neutral gas measurements for the remaining cases.
This study has led us to find that:

\begin{enumerate}
\item The \cf\ absorption line profiles are complex, showing 
  both narrow ($b\!<\!10$\,\kms) and broad ($b\!>\!10$\,\kms)
  components, which trace cool, photoionized and
  hot, collisionally ionized gas, respectively (we used \os\ to study
  the hot gas in Paper I).

\item The median \cf\ column densities in DLAs and sub-DLAs 
  [$\langle$log$N_{\rm \cf}$(DLA)$\rangle$=14.2] 
  are substantially higher than typical IGM values
  [$\langle$log$N_{\rm \cf}$(IGM)$\rangle$=13.0], 
  but similar to those seen in LBGs
  [$\langle$log$N_{\rm \cf}$(LBG)$\rangle$=14.0].

\item The total \cf\ line width is broader than the total neutral line width
  in 69 of 74 cases. 

\item The total \cf\ column density, line width, and velocity offset
  in the DLAs and sub-DLAs are all correlated, so the strongest \cf\
  absorbers also tend to be the broadest and most offset relative to
  the neutral gas.

\item The total \cf\ column density is correlated with the
  neutral-phase metallicity. The significance of this correlation is
  $>$6.0$\sigma$, 
  and it is found even when the saturated cases, the proximate DLAs,
  and the sub-DLAs are excluded.

\item The total \cf\ line width is weakly correlated with the
  neutral-phase metallicity. This correlation is detected at
  $4.1\sigma$ significance (using $(v_+-v_-)_{\rm \cf}$ 
  to measure the total line width), 
  and is found independently in the lower-
  and upper- redshift halves of the sample. The slope of the
  metallicity/high-ion line width relation is similar to the slope of the
  metallicity/low-ion line width relation reported by \citet{Le06}.

\item None of the \cf\ properties (column density, line width, central
  velocity) correlate with the \hi\ column density, even though our
  sample spans a factor of 100 in $N_{\rm \hi}$. 
  Indeed, though we only have \nsub\ sub-DLAs against \ndla\ DLAs, the
  mean values of log\,$N$, $\Delta v$, and $|\bar{v}|$ are the same in DLAs
  as in sub-DLAs.
  However, assuming a constant ionization correction 
  $N_{\rm \cf}/N_{\rm C}=0.3$, then the sub-DLAs show a mean 
  $N_{\rm \hw}$ of 19.36, whereas the
  DLAs show a mean $N_{\rm \hw}$ of 19.77, a factor of $\approx$2.5 higher.
  This is because sub-DLAs show (on average) similar \cf\ columns but
  higher metallicities.
 
\item We find slighty lower mean \cf\ column densities and total line
  widths among the seven proximate DLAs/sub-DLAs than among the 67
  intervening systems. This trend is worth investigating using a
  larger proximate sample.

\item The mean velocity offset between the \cf\ and the neutral gas
  $|\bar{v}|_{\rm \cf}$ 
  has a mean value of 69\,\kms\ over our \ntot\ DLAs and 
  sub-DLAs, implying a net amount of outflow or inflow is present. 
  The maximum observed \cf\ velocity $v_{\rm max}$ reaches
  $>$200\,\kms\ in 42/\ntot\ cases, and $>$500\,\kms\ in eight cases.
  $v_{\rm max}$ is correlated with
  the metallicity at the 2.9$\sigma$ level. 
 
\item We calculate the escape velocity from the width of the neutral
  line absorption. We observe \cf\ moving above the escape velocity in
  25 DLAs and 4 sub-DLAs, covering 2.5 orders of magnitude of 
  [Z/H]. In other words, \cf\ clouds that are unbound from the central
  potential well are seen in $\approx$40\% of DLAs and sub-DLAs. 
  Assuming a characteristic \cf\ radius of 40\,kpc,
  these wind candidate absorbers show typical (median) masses of
  $\sim2\times10^9$\,M$_\odot$, 
  kinetic energies of $\sim1\times10^{57}$\,erg, mass flow rates of
  $\sim22$\,\mfu, and kinetic energy injection rates 
  $\dot{E}_k\sim4\times10^{41}$\,erg\,s$^{-1}$. 
  The typical value for $\dot{E}_k$ requires a
  SFR per DLA of $\sim$2\,\mfu, or a SFR per unit area of
  $\sim3\times10^{-4}$\,\mfu\,kpc$^{-2}$, to power the winds.

\end{enumerate}

We conclude with several remarks concerning the origin of ionized gas in DLAs
and sub-DLAs. Since $\Delta v_{\rm \cf}$ is almost always broader than 
the gravitationally broadened $\Delta v_{\rm Neut}$, an additional
energy source is required to heat and accelerate the \cf-bearing
clouds, in $\approx$40\% of cases to above to escape speed.
We propose that star formation and supernovae can provide this source.
DLA/sub-DLA galaxies with higher rates of star formation will 
produce higher EUV fluxes from massive stars, photoionizing the gas
that is seen in the narrow \cf\ components, and will also undergo higher rates
of Type II supernovae and stellar winds. The supernovae lead to 
(i) metal enrichment through nucleosynthesis, 
(ii) the generation of million-degree interstellar plasma, which can
interact with embedded clouds to form gas containing \cf\ and \os, and
(iii) the injection of mechanical energy to the surrounding
ISM, explaining the extended velocity fields for the ionized gas. 
Such a scenario would explain (at least qualitatively) the
metallicity-\cf\ line width correlation, and the velocity and
ionization level of the wind candidates.

Infalling clouds can also contribute to the \cf\ seen in DLAs
\citep[e.g.][]{WP00b}. However, as we have pointed out, infall cannot
explain the highest velocity \cf\ components, which are detected at
well over the escape speed. Although metallicity measurements could in theory
discriminate between the infall and outflow hypotheses, it is very
difficult to directly measure the metallicity of the ionized gas in DLAs.
However, one practical test of the idea that star formation leads to the
production of ionized gas in DLAs (and the associated idea that the
high-velocity \cf\ components in DLAs trace winds) would be a detailed
comparison between the properties of \cf\ absorption in DLAs and in
LBGs, where outflows are \emph{directly} observed \citep{Pe00, Pe02,
  Sh03}. Though we have not yet conducted a full comparison, we do
note that the similar \cf\ column density distributions observed in
DLAs and LBGs suggest that the ionized gas in these two classes of
object shares a common origin in supernova-driven outflows. 

\vskip 1cm
{\bf Acknowledgements}\\
AJF gratefully acknowledges the support of a Marie Curie
Intra-European Fellowship awarded by the European Union Sixth
Framework Programme. 
PP and RS gratefully acknowledge support from the Indo-French Centre
for the Promotion of Advanced Research (Centre Franco-Indien pour la
Promotion de la Recherche Avanc\'ee) under contract No. 3004-3.
We thank Alain Smette for kindly providing the spectra of many quasars
in the Hamburg-ESO survey prior to publication. We acknowledge
valuable discussions with Blair Savage, Jason Prochaska, and Art
Wolfe. Finally, we thank the referee for perceptive comments that
improved the paper.

\input{tab1.tex}
\input{tab2.tex}
\input{tab3.tex}

\setcounter{figure}{0}
\begin{figure*}
\includegraphics[width=17cm]{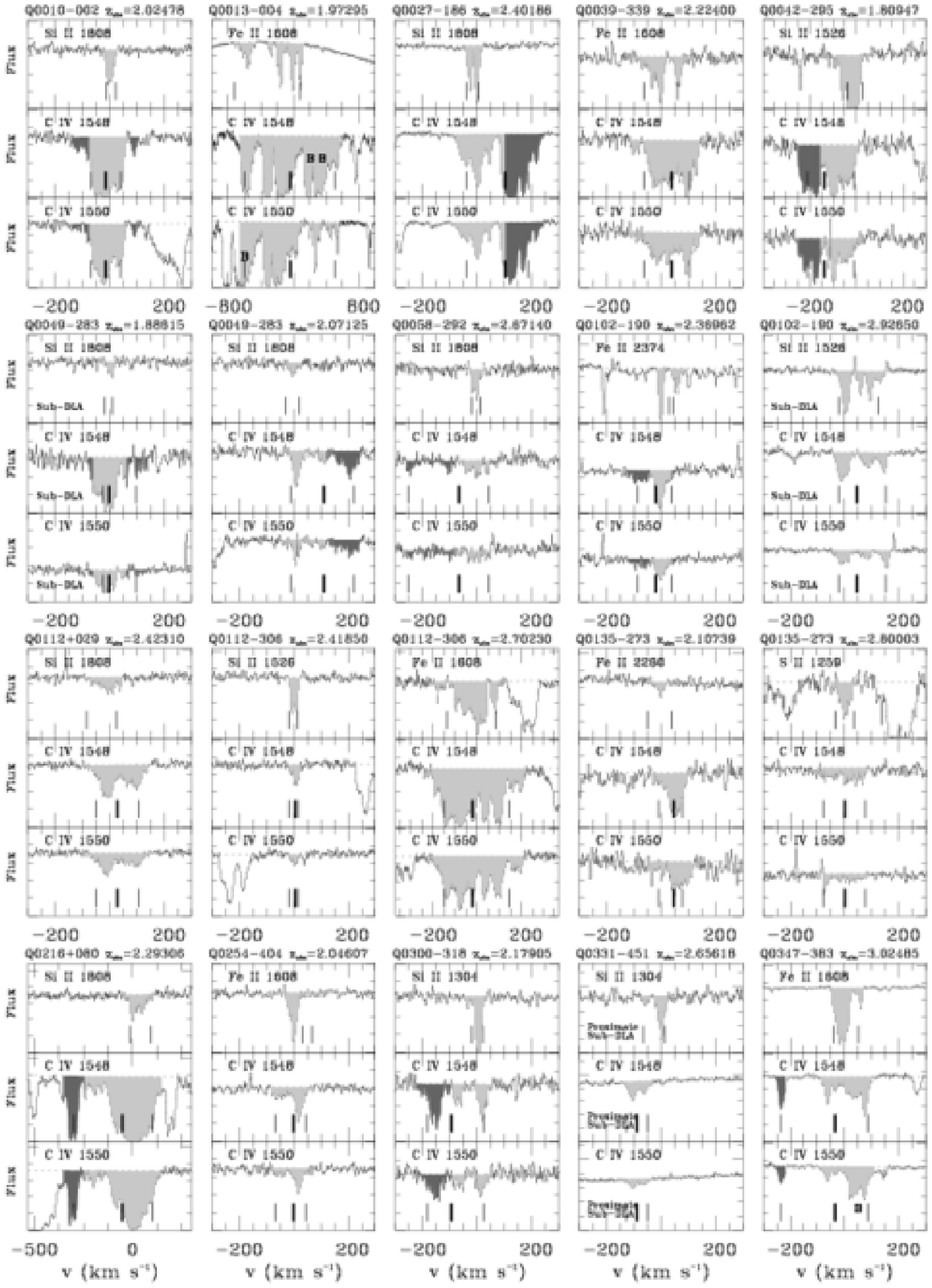}
\caption{
  VLT/UVES absorption line spectra of \cf\ $\lambda\lambda1548, 1550$ 
  and an optically thin line (typically \ion{Si}{ii} or \ion{Fe}{ii}) 
  chosen to trace the neutral phase component structure, for all DLAs and
  sub-DLAs in our sample. 
  The flux is in arbitrary units, with the bottom of each panel at zero.
  The light (dark) shaded regions show \cf\ absorption
  below (above) $v_{\rm esc}$, where $v_{\rm esc}=2.4\Delta v_{\rm
  Neut}$ (see text).
  The thick vertical line in each \cf\ panel denotes the optical-depth
  weighted mean velocity of the profile. 
  The two narrow vertical lines show the velocities corresponding to 
  5\% and 95\% of the integrated optical depth; the interval between
  these velocities defines the total line width $\Delta v$.
  A small letter ``B'' within the shaded area indicates a
  blend; in these cases the other \cf\ line was used for measurement.
  The label ``Proximate'' implies the absorber is at $<$5\,000\,\kms\
  from the QSO. Note how the wind candidate absorbers (the dark
  regions) show no absorption in the weak neutral line shown here,
  implying these absorbers have low column densities of neutral gas
  and are highly ionized.
  } 
\end{figure*}
\addtocounter{figure}{-1}
\begin{figure*}
\includegraphics[width=17cm]{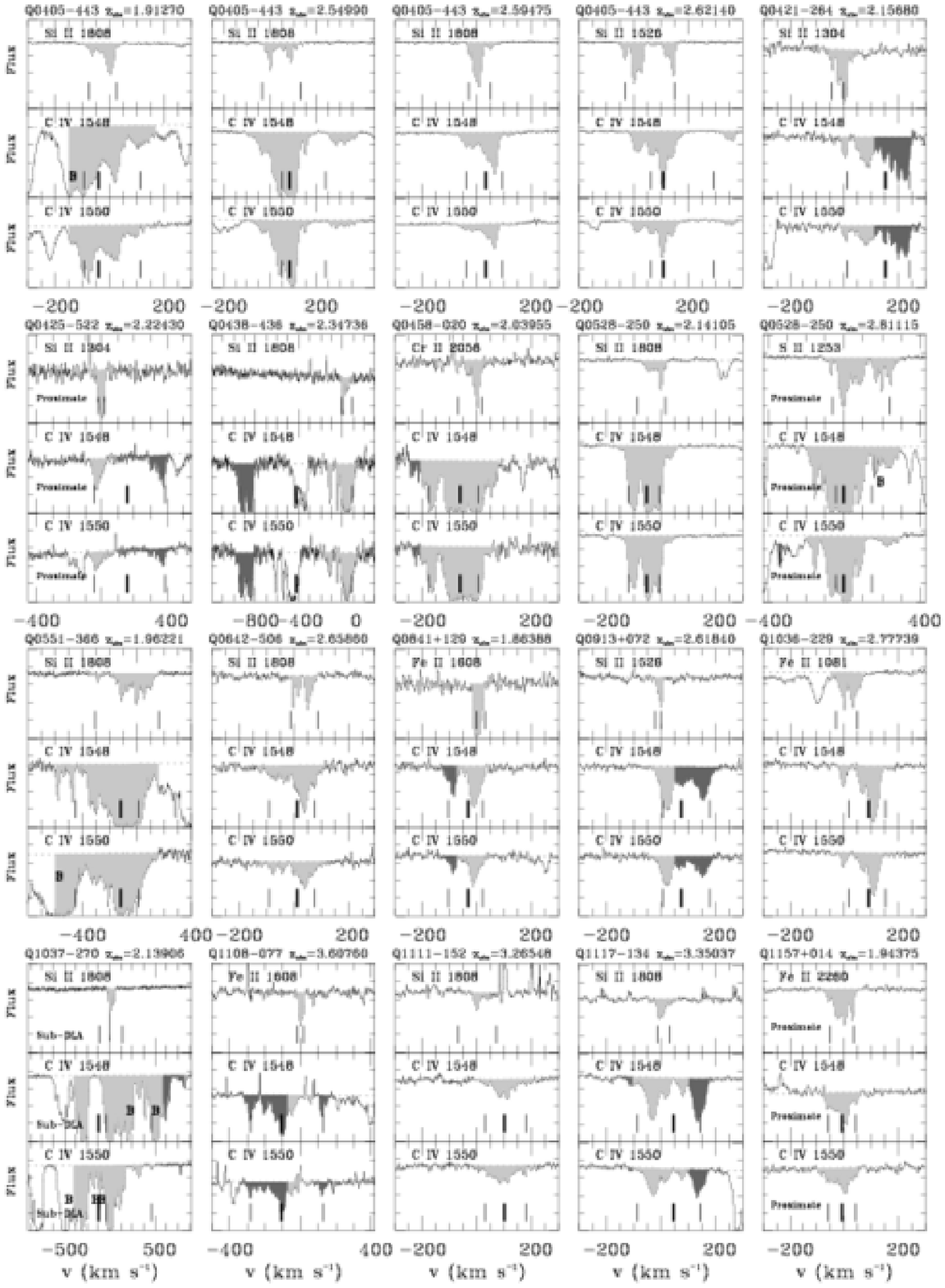}
\caption{(-continued).}
\end{figure*}
\addtocounter{figure}{-1}
\begin{figure*}
\includegraphics[width=17cm]{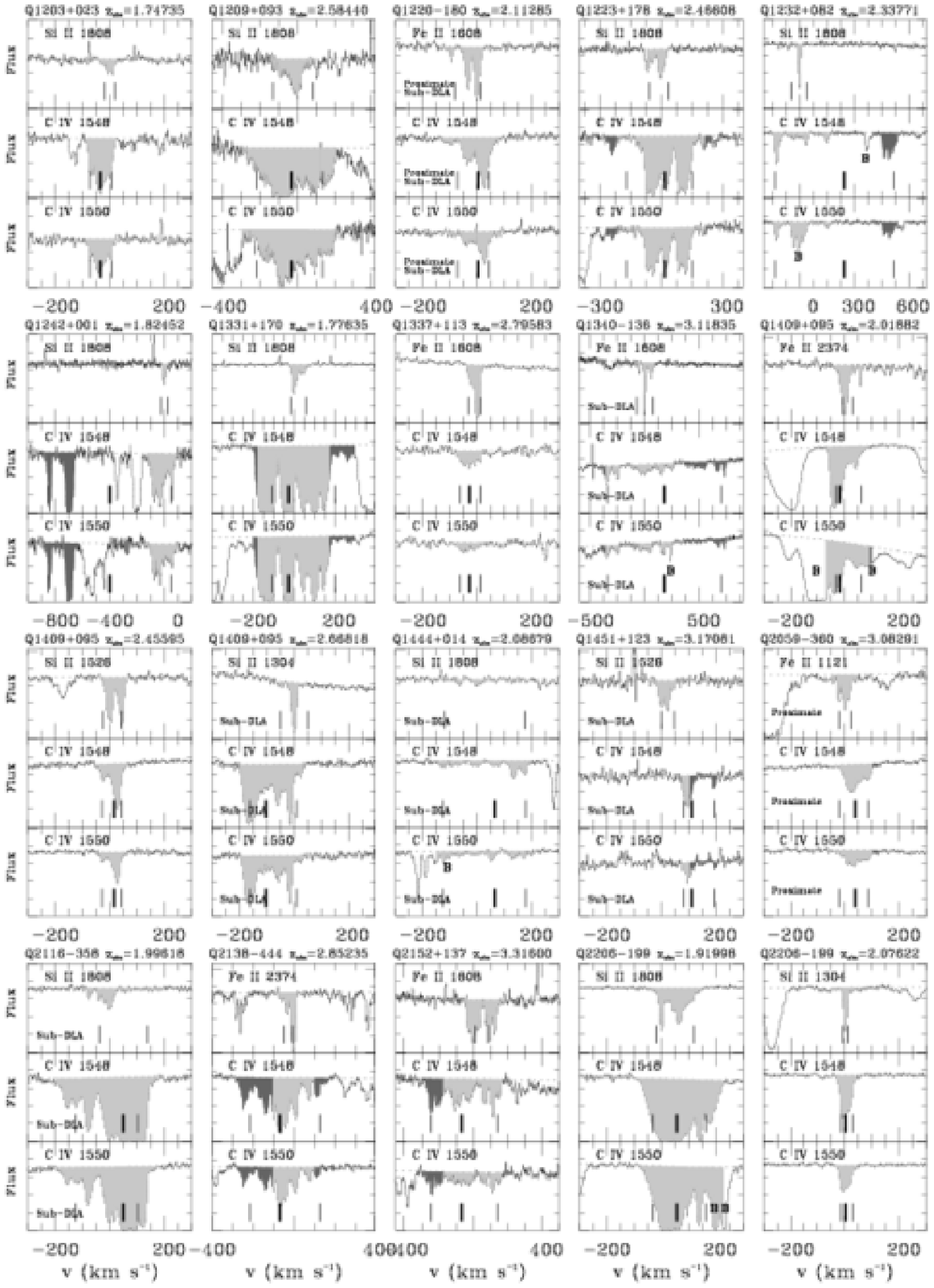}
\caption{(-continued).}
\end{figure*}
\addtocounter{figure}{-1}
\begin{figure*}
\includegraphics[width=17cm]{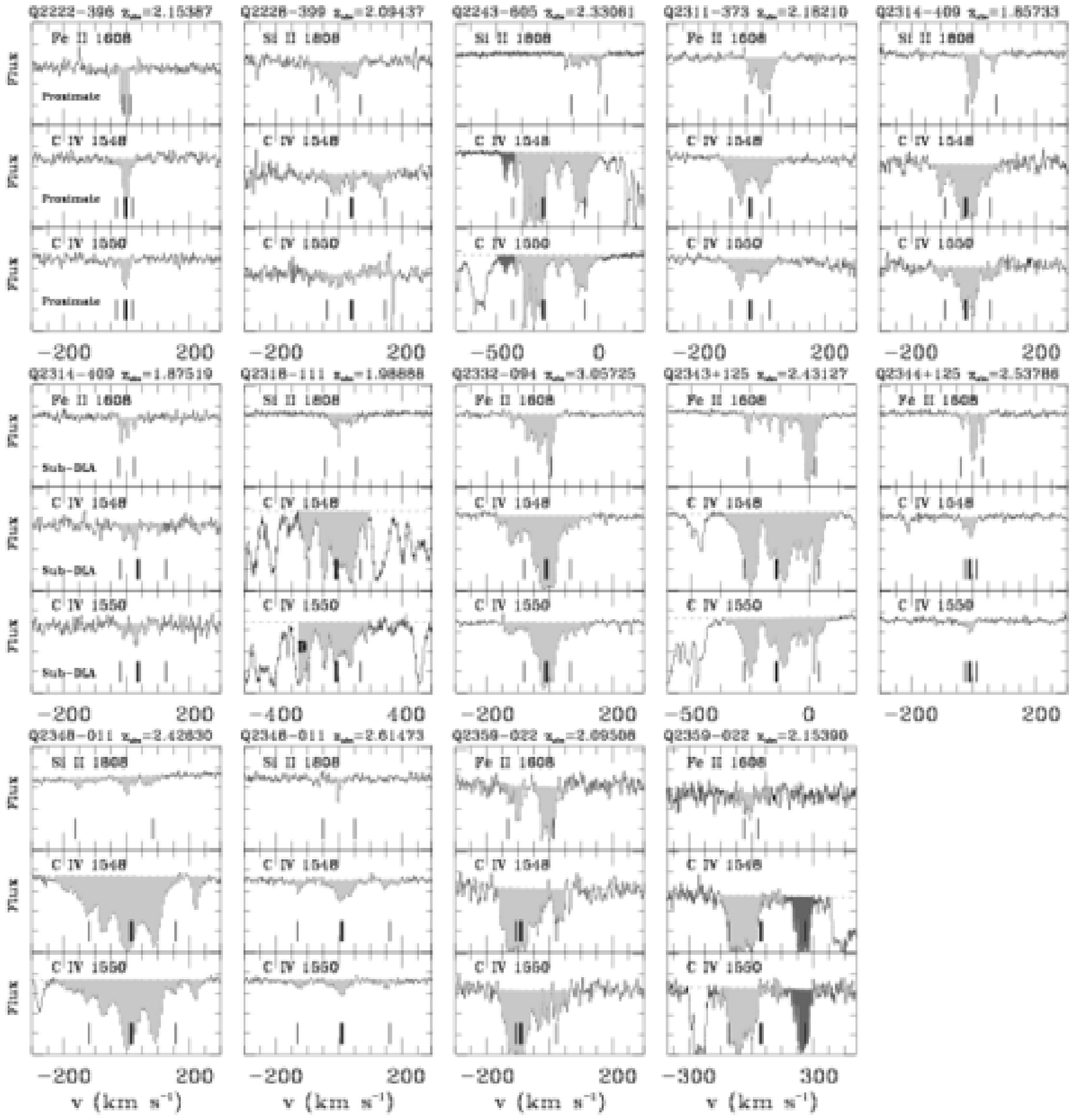}
\caption{(-continued).}
\end{figure*}

\end{document}

%% file: tab1.tex
\begin{longtable}{lcccc ccccc c}
\caption{Comparison of \cf\ and Neutral DLA Measurements}\\
\hline\hline
QSO & $z_{\rm em}$ & $z_{\rm abs}$$^1$ & \multicolumn{4}{c}{\underline{~~~~~~~~~~~Neutral Gas Properties$^2$~~~~~~~~~~~}} &\multicolumn{4}{c}{\underline{~~~~~~~~~~~~~~~~~~~\cf~Properties~~~~~~~~~~~~~~~~~~~~}}\\
 & & & log $N_{\rm \hi}$ & [Z/H] & Z & $\Delta v_{\rm Neut}$$^3$ & Line$^4$ & log$N_{\rm \cf}$$^5$ & $\Delta v_{\rm \cf}$$^6$ &  $\bar{v}_{\rm \cf}$$^7$\\
\hline
\endfirsthead
\caption{continued.}\\
\hline\hline
QSO & $z_{\rm em}$ & $z_{\rm abs}$$^1$ & \multicolumn{4}{c}{\underline{~~~~~~~~~~~Neutral Gas Properties$^2$~~~~~~~~~~~}} &\multicolumn{4}{c}{\underline{~~~~~~~~~~~~~~~~~~~\cf~Properties~~~~~~~~~~~~~~~~~~~~}}\\
 & & & log $N_{\rm \hi}$ & [Z/H] & Z & $\Delta v_{\rm Neut}$$^3$ & Line$^4$ & log$N_{\rm \cf}$$^5$ & $\Delta v_{\rm \cf}$$^6$ &  $\bar{v}_{\rm \cf}$$^7$\\
\hline \endhead \hline \endfoot
               Q0010--002 & 2.15 & 2.02478 & 20.95$\pm$0.10 &     $-$1.43$\pm$0.11 & Zn & 32 &       1550 &        $>$14.65 &         $<$110 &    $-$15$\pm$6 \\
               Q0013--004 & 2.09 & 1.97295 & 20.83$\pm$0.05 &    $-$0.59$\pm$0.05 & Zn & 720 &   1550$^8$ &        $>$15.41 &        $<$1111 &   $-$42$\pm$40 \\
               Q0027--186 & 2.56 & 2.40186 & 21.75$\pm$0.10 &     $-$1.63$\pm$0.10 & Zn & 44 &       1550 &        $>$14.68 &         $<$227 &      104$\pm$6 \\
               Q0039--339 & 2.48 & 2.22400 & 20.60$\pm$0.10 &    $-$1.31$\pm$0.12 & Si & 122 &       1550 &  14.41$\pm$0.01 &      168$\pm$3 &       39$\pm$3 \\
               Q0042--295 & 2.39 & 1.80947 & 20.40$\pm$0.10 &     $-$1.25$\pm$0.15 & Si & 65 &       1550 &        $>$14.65 &         $<$200 &   $-$137$\pm$8 \\
               Q0049--283 & 2.26 & 1.88615 & 20.20$\pm$0.08 &      $-$1.03$\pm$0.09 & S & 26 &       1550 &  14.25$\pm$0.02 &      123$\pm$4 &     $-$3$\pm$3 \\
               Q0049--283 & 2.26 & 2.07125 & 20.45$\pm$0.10 &     $-$1.31$\pm$0.12 & Si & 51 &       1548 &  13.83$\pm$0.02 &      229$\pm$3 &      110$\pm$5 \\
               Q0058--292 & 3.09 & 2.67140 & 21.10$\pm$0.10 &     $-$1.53$\pm$0.10 & Zn & 34 &       1548 &  13.54$\pm$0.05 &     293$\pm$13 &   $-$68$\pm$17 \\
               Q0102--190 & 3.04 & 2.36962 & 21.00$\pm$0.08 &      $-$1.90$\pm$0.08 & S & 17 &       1550 &  13.91$\pm$0.02 &      123$\pm$3 &    $-$19$\pm$3 \\
               Q0102--190 & 3.04 & 2.92650 & 20.00$\pm$0.10 &    $-$1.50$\pm$0.10 & Si & 146 &       1548 &  13.75$\pm$0.01 &      172$\pm$3 &       47$\pm$3 \\
                Q0112+029 & 2.81 & 2.42310 & 20.90$\pm$0.10 &     $-$1.31$\pm$0.11 & S & 112 &       1548 &  13.95$\pm$0.01 &      161$\pm$3 &       27$\pm$3 \\
               Q0112--306 & 2.99 & 2.41850 & 20.50$\pm$0.08 &     $-$2.42$\pm$0.08 & Si & 31 &       1548 &  13.10$\pm$0.02 &       35$\pm$3 &        6$\pm$3 \\
               Q0112--306 & 2.99 & 2.70230 & 20.30$\pm$0.10 &    $-$0.49$\pm$0.11 & Si & 218 &       1550 &        $>$14.83 &         $<$295 &    $-$22$\pm$6 \\
               Q0135--273 & 3.21 & 2.10739 & 20.30$\pm$0.15 &      $-$1.12$\pm$0.16 & S & 92 &       1548 &  13.94$\pm$0.01 &       88$\pm$3 &       47$\pm$3 \\
               Q0135--273 & 3.21 & 2.80003 & 21.00$\pm$0.10 &      $-$1.40$\pm$0.10 & S & 66 &       1548 &  13.56$\pm$0.02 &      153$\pm$3 &        0$\pm$3 \\
                Q0216+080 & 2.99 & 2.29306 & 20.50$\pm$0.10 &    $-$0.70$\pm$0.11 & Zn & 104 &       1550 &        $>$15.09 &         $<$399 &   $-$51$\pm$12 \\
               Q0254--404 & 2.28 & 2.04607 & 20.45$\pm$0.08 &      $-$1.55$\pm$0.09 & S & 37 &       1548 &  13.69$\pm$0.01 &      111$\pm$3 &        0$\pm$3 \\
               Q0300--318 & 2.37 & 2.17905 & 20.80$\pm$0.10 &      $-$1.80$\pm$0.10 & S & 41 &       1548 &  14.03$\pm$0.02 &      211$\pm$3 &    $-$96$\pm$4 \\
               Q0331--451 & 2.67 & 2.65618 & 19.82$\pm$0.05 &     $-$1.49$\pm$0.05 & Si & 80 &       1548 &  13.35$\pm$0.01 &       81$\pm$3 &    $-$88$\pm$3 \\
               Q0347--383 & 3.22 & 3.02485 & 20.73$\pm$0.05 &     $-$1.17$\pm$0.07 & Zn & 93 &       1548 &  13.95$\pm$0.01 &      319$\pm$3 &    $-$35$\pm$3 \\
               Q0405--443 & 3.02 & 1.91270 & 20.80$\pm$0.10 &     $-$1.03$\pm$0.10 & Zn & 98 &       1550 &        $>$14.60 &         $<$205 &    $-$42$\pm$6 \\
               Q0405--443 & 3.02 & 2.54990 & 21.15$\pm$0.15 &    $-$1.36$\pm$0.16 & Zn & 165 &       1550 &        $>$14.77 &         $<$187 &      83$\pm$12 \\
               Q0405--443 & 3.02 & 2.59475 & 21.05$\pm$0.10 &     $-$1.12$\pm$0.10 & Zn & 79 &       1548 &  13.86$\pm$0.01 &      131$\pm$3 &       31$\pm$3 \\
               Q0405--443 & 3.02 & 2.62140 & 20.45$\pm$0.10 &    $-$2.04$\pm$0.10 & Si & 182 &       1548 &  14.06$\pm$0.01 &      231$\pm$3 &      109$\pm$3 \\
               Q0421--264 & 2.28 & 2.15680 & 20.65$\pm$0.10 &     $-$1.86$\pm$0.10 & Si & 47 &       1548 &  14.18$\pm$0.01 &      227$\pm$3 &      152$\pm$3 \\
               Q0425--522 & 2.25 & 2.22430 & 20.30$\pm$0.10 &      $-$1.43$\pm$0.11 & S & 40 &       1548 &  14.00$\pm$0.01 &      435$\pm$3 &      156$\pm$5 \\
               Q0438--436 & 2.86 & 2.34736 & 20.78$\pm$0.12 &     $-$0.72$\pm$0.13 & Zn & 89 &   1550$^8$ &        $>$15.03 &        $<$1035 &  $-$425$\pm$40 \\
               Q0458--020 & 2.29 & 2.03955 & 21.70$\pm$0.10 &     $-$1.22$\pm$0.10 & Zn & 88 &       1550 &        $>$15.06 &         $<$185 &   $-$63$\pm$46 \\
               Q0528--250 & 2.77 & 2.14105 & 20.98$\pm$0.05 &    $-$1.36$\pm$0.06 & Zn & 105 &       1550 &        $>$14.70 &         $<$113 &   $-$51$\pm$12 \\
               Q0528--250 & 2.77 & 2.81115 & 21.35$\pm$0.07 &    $-$0.91$\pm$0.07 & Zn & 304 &       1550 &        $>$15.09 &         $<$188 &    $-$4$\pm$12 \\
               Q0551--366 & 2.32 & 1.96221 & 20.70$\pm$0.08 &    $-$0.35$\pm$0.08 & Zn & 468 &       1548 &        $>$15.16 &         $<$461 &  $-$121$\pm$38 \\
               Q0642--506 & 3.09 & 2.65860 & 20.95$\pm$0.08 &     $-$1.05$\pm$0.09 & Zn & 99 &       1548 &  14.00$\pm$0.01 &      168$\pm$3 &       15$\pm$3 \\
                Q0841+129 & 2.50 & 1.86388 & 21.00$\pm$0.10 &      $-$1.51$\pm$0.11 & S & 32 &       1548 &  13.89$\pm$0.01 &      126$\pm$3 &    $-$33$\pm$3 \\
                Q0913+072 & 2.78 & 2.61840 & 20.35$\pm$0.10 &     $-$2.59$\pm$0.10 & Si & 22 &       1548 &  14.13$\pm$0.01 &      175$\pm$3 &       76$\pm$3 \\
               Q1036--229 & 3.13 & 2.77739 & 20.93$\pm$0.05 &      $-$1.36$\pm$0.05 & S & 80 &       1550 &  14.11$\pm$0.01 &      133$\pm$3 &       90$\pm$3 \\
               Q1037--270 & 2.20 & 2.13906 & 19.70$\pm$0.05 &    $-$0.31$\pm$0.05 & Zn & 250 &       1550 &        $>$15.32 &         $<$496 &  $-$125$\pm$12 \\
               Q1108--077 & 3.92 & 3.60760 & 20.37$\pm$0.07 &     $-$1.59$\pm$0.07 & Si & 31 &       1548 &  14.34$\pm$0.01 &      410$\pm$3 &   $-$108$\pm$3 \\
               Q1111--152 & 3.37 & 3.26548 & 21.30$\pm$0.05 &    $-$1.65$\pm$0.11 & Zn & 140 &       1548 &  13.54$\pm$0.01 &      149$\pm$3 &      101$\pm$3 \\
               Q1117--134 & 3.96 & 3.35037 & 20.95$\pm$0.10 &     $-$1.41$\pm$0.11 & Zn & 44 &       1548 &  14.15$\pm$0.01 &      233$\pm$3 &       46$\pm$3 \\
                Q1157+014 & 1.99 & 1.94375 & 21.80$\pm$0.10 &     $-$1.44$\pm$0.10 & Zn & 89 &       1548 &  13.86$\pm$0.01 &      102$\pm$3 &     $-$9$\pm$3 \\
                Q1203+023 & 2.13 & 1.74735 & 20.40$\pm$0.10 &     $-$0.97$\pm$0.11 & Zn & 38 &       1550 &  14.17$\pm$0.01 &       77$\pm$3 &    $-$36$\pm$3 \\
                Q1209+093 & 3.30 & 2.58440 & 21.40$\pm$0.10 &    $-$1.01$\pm$0.10 & Zn & 214 &       1550 &        $>$15.03 &         $<$355 &   $-$34$\pm$20 \\
               Q1220--180 & 2.16 & 2.11285 & 20.12$\pm$0.07 &      $-$0.93$\pm$0.07 & S & 95 &       1548 &  14.01$\pm$0.01 &      119$\pm$3 &        3$\pm$3 \\
                Q1223+178 & 2.94 & 2.46608 & 21.40$\pm$0.10 &     $-$1.63$\pm$0.10 & Zn & 91 &       1550 &  14.71$\pm$0.01 &      320$\pm$3 &       18$\pm$3 \\
                Q1232+082 & 2.57 & 2.33771 & 20.90$\pm$0.08 &      $-$1.43$\pm$0.08 & S & 85 &       1548 &  13.93$\pm$0.01 &      653$\pm$3 &      246$\pm$4 \\
                Q1242+001 & 2.08 & 1.82452 & 20.45$\pm$0.10 &     $-$1.18$\pm$0.12 & Zn & 56 &   1550$^8$ &        $>$14.76 &         $<$900 &  $-$400$\pm$40 \\
                Q1331+170 & 2.08 & 1.77635 & 21.15$\pm$0.07 &     $-$1.28$\pm$0.08 & Zn & 75 &       1550 &        $>$15.27 &         $<$308 &   $-$24$\pm$12 \\
                Q1337+113 & 2.92 & 2.79583 & 21.00$\pm$0.08 &     $-$1.86$\pm$0.09 & Si & 42 &       1548 &  13.36$\pm$0.03 &       78$\pm$4 &    $-$29$\pm$4 \\
               Q1340--136 & 3.20 & 3.11835 & 20.05$\pm$0.08 &     $-$1.42$\pm$0.08 & S & 153 &   1550$^8$ &  14.05$\pm$0.05 &    1044$\pm$20 &     180$\pm$20 \\
                Q1409+095 & 2.85 & 2.01882 & 20.65$\pm$0.10 &     $-$1.62$\pm$0.16 & Zn & 39 &       1548 &        $>$14.15 &         $<$ 93 &    $-$20$\pm$6 \\
                Q1409+095 & 2.85 & 2.45595 & 20.53$\pm$0.08 &     $-$2.06$\pm$0.08 & Si & 69 &       1548 &  13.74$\pm$0.01 &       74$\pm$3 &       15$\pm$3 \\
                Q1409+095 & 2.85 & 2.66818 & 19.80$\pm$0.08 &     $-$1.41$\pm$0.09 & S & 100 &       1550 &  14.45$\pm$0.01 &      172$\pm$3 &   $-$102$\pm$3 \\
                Q1444+014 & 2.21 & 2.08679 & 20.25$\pm$0.07 &    $-$0.80$\pm$0.09 & Zn & 294 &       1548 &  13.52$\pm$0.02 &      305$\pm$5 &       63$\pm$5 \\
                Q1451+123 & 3.25 & 3.17081 & 20.20$\pm$0.20 &     $-$2.10$\pm$0.21 & Si & 45 &       1548 &  13.55$\pm$0.03 &      115$\pm$5 &     112$\pm$11 \\
               Q2059--360 & 3.09 & 3.08291 & 20.98$\pm$0.08 &      $-$1.77$\pm$0.09 & S & 44 &       1548 &  13.73$\pm$0.01 &      103$\pm$3 &       39$\pm$3 \\
               Q2116--358 & 2.34 & 1.99618 & 20.10$\pm$0.07 &    $-$0.34$\pm$0.11 & Zn & 177 &       1550 &        $>$15.09 &         $<$231 &      49$\pm$12 \\
               Q2138--444 & 3.17 & 2.85235 & 20.98$\pm$0.05 &     $-$1.74$\pm$0.05 & Zn & 60 &       1548 &  14.45$\pm$0.01 &      342$\pm$3 &    $-$67$\pm$3 \\
                Q2152+137 & 4.26 & 3.31600 & 20.50$\pm$0.15 &     $-$1.37$\pm$0.15 & Si & 74 &       1548 &  14.29$\pm$0.01 &      389$\pm$3 &    $-$66$\pm$3 \\
               Q2206--199 & 2.56 & 1.91998 & 20.67$\pm$0.05 &    $-$0.54$\pm$0.05 & Zn & 136 &       1548 &        $>$14.96 &         $<$195 &      57$\pm$12 \\
               Q2206--199 & 2.56 & 2.07622 & 20.44$\pm$0.05 &     $-$2.32$\pm$0.05 & Si & 20 &       1548 &  13.71$\pm$0.01 &       47$\pm$3 &        3$\pm$3 \\
               Q2222--396 & 2.18 & 2.15387 & 20.85$\pm$0.10 &      $-$1.97$\pm$0.10 & S & 21 &       1548 &  13.44$\pm$0.02 &       49$\pm$3 &     $-$3$\pm$3 \\
               Q2228--399 & 2.21 & 2.09437 & 21.20$\pm$0.10 &    $-$1.36$\pm$0.12 & Zn & 138 &       1548 &  13.70$\pm$0.02 &      186$\pm$5 &       43$\pm$5 \\
               Q2243--605 & 3.01 & 2.33061 & 20.65$\pm$0.05 &    $-$0.85$\pm$0.05 & Zn & 173 &       1550 &        $>$14.73 &         $<$349 &  $-$272$\pm$14 \\
               Q2311--373 & 2.48 & 2.18210 & 20.48$\pm$0.13 &             $<-$1.33 & Zn & 77 &       1548 &  13.95$\pm$0.01 &      127$\pm$3 &    $-$36$\pm$3 \\
               Q2314--409 & 2.45 & 1.85733 & 20.90$\pm$0.10 &     $-$1.02$\pm$0.14 & Zn & 95 &       1548 &  14.24$\pm$0.01 &      143$\pm$3 &    $-$24$\pm$3 \\
               Q2314--409 & 2.45 & 1.87519 & 20.10$\pm$0.20 &             $<-$1.19 & Zn & 49 &       1548 &  13.38$\pm$0.02 &      140$\pm$4 &       37$\pm$4 \\
               Q2318--111 & 2.96 & 1.98888 & 20.68$\pm$0.05 &    $-$0.85$\pm$0.06 & Zn & 207 &       1548 &  14.68$\pm$0.01 &      329$\pm$3 &    $-$11$\pm$3 \\
               Q2332--094 & 3.32 & 3.05725 & 20.50$\pm$0.07 &     $-$1.33$\pm$0.08 & S & 111 &       1550 &        $>$14.54 &         $<$150 &    $-$11$\pm$6 \\
                Q2343+125 & 2.51 & 2.43127 & 20.40$\pm$0.07 &    $-$0.87$\pm$0.07 & Zn & 289 &       1550 &  14.69$\pm$0.01 &      315$\pm$3 &   $-$137$\pm$3 \\
                Q2344+125 & 2.76 & 2.53786 & 20.50$\pm$0.10 &     $-$1.81$\pm$0.10 & Si & 69 &       1548 &  13.02$\pm$0.02 &       40$\pm$3 &    $-$12$\pm$3 \\
               Q2348--011 & 3.01 & 2.42630 & 20.50$\pm$0.10 &     $-$0.62$\pm$0.10 & S & 248 &       1550 &        $>$14.77 &         $<$277 &       15$\pm$6 \\
               Q2348--011 & 3.01 & 2.61473 & 21.30$\pm$0.08 &    $-$2.02$\pm$0.08 & Si & 100 &       1548 &  13.53$\pm$0.01 &      296$\pm$4 &       11$\pm$3 \\
               Q2359--022 & 2.81 & 2.09508 & 20.65$\pm$0.10 &    $-$0.84$\pm$0.13 & Zn & 146 &       1550 &        $>$14.87 &         $<$131 &   $-$92$\pm$66 \\
               Q2359--022 & 2.81 & 2.15390 & 20.30$\pm$0.10 &     $-$1.62$\pm$0.10 & Si & 67 &       1550 &        $>$15.01 &         $<$367 &      48$\pm$42 \\
\end{longtable}
$^1$ The absorber redshift is defined by the velocity of the strongest component seen in the low-ionization lines.\\
$^2$ Taken from \citet{Le06}, \citet{Sm05}, Smette et al. (2007, in preparation), \citet{Ak05}, \cite{EL01}, or this paper. The element Z is Zn if \ion{Zn}{ii} is detected, otherwise Si or S.\\
$^3$ Neutral line velocity width containing central 90\% of the apparent optical depth, in \kms. The measurement is made on an optically thin line, i.e. one with $0.1<F(v_0)/F_c(v_0)<0.6$ \citep[see][]{Le06}. Typical error is $\approx3$\,\kms.\\
$^4$ Line used to measure \cf. $\lambda$1548 is chosen if $F(v_0)/F_c(v_0)>0.1$, $\lambda$1550 otherwise.\\
$^5$ \cf\ column density measured using the apparent optical depth method. $N_{\rm \cf}$ is in \sqcm. Lower limits are 3$\sigma$.\\
$^6$ \cf\ line width containing central 90\% of apparent optical depth, in \kms. Upper limits represent saturated cases.\\
$^7$ Mean velocity of \cf\ absorption profile, i.e. velocity offset from neutral gas, in \kms. Errors have been doubled for saturated cases.\\
$^8$ Both lines partly blended; results from combining measurements of absorption in two unblended velocity ranges.\\

%% file: tab2.tex
\begin{table}
\begin{minipage}[t]{\columnwidth}
\caption{Statistical Significance of Correlations: Kendall $\tau$ Analysis}
\renewcommand{\footnoterule}{}
\begin{tabular}{lllll l}
\hline\hline
Quan. 1 & Quan. 2 & Sample & Size & $\tau$ & $\sigma$\\
\hline
                                                 $N_{\rm \cf}$ & $(v_+-v_-)_{\rm \cf}$ & all & 74 & 0.54 & $>$6.0 \\
                                                                                                           &  & interv DLA\footnote{In the intervening DLA sample, sub-DLAs and DLAs at $<$5\,000\,\kms\ from the QSO were excluded.} & 58  & 0.52 & $>$6.0 \\
                                                           &  & unsaturated\footnote{In the unsaturated sample (considered for correlations involving $N_{\rm \cf}$ or $\Delta v_{\rm \cf}$), all saturated \cf\ lines were removed.} & 49  & 0.45 & 4.5 \\
\hline
                                        $|\bar{v}|_{\rm \cf}$ & $(v_+-v_-)_{\rm \cf}$ & all  & 74 & 0.34 & 4.3 \\
                                                                            &  & interv DLA  & 58 & 0.35 & 3.8 \\
\hline
                                                   ${\rm [Z/H]}$ & $N_{\rm \cf}$ & all & 74 &       0.45 & $>$6.0 \\
                                                                                                                                                                                                                      &  & interv DLA & 58  & 0.46 & 5.1 \\
                                                                  &  & \ion{Zn}{ii}\footnote{In the \ion{Zn}{ii} sample, cases where [Z/H] was derived from Si or S were excluded. This only applies to correlations involving [Z/H].} & 37 & 0.40 & 3.5 \\
                                                                                                                                                                                                                     &  & unsaturated & 49  & 0.25 & 2.5 \\
\hline
                 ${\rm [Z/H]}$ & $\Delta v_{\rm \cf}$ & all  & 74 & 0.28 & 3.4 \\
               &  & interv DLA & 58 & 0.27 & 2.9 \\
             &  & \ion{Zn}{ii} & 37 & 0.30 & 2.6 \\
                                                                                                                                                                                                                     &  & unsaturated & 49  & 0.14 & 1.4 \\
\hline
                ${\rm [Z/H]}$ & $(v_+-v_-)_{\rm \cf}$ & all  & 74 & 0.33 & 4.1 \\
               &  & interv DLA & 58 & 0.30 & 3.3 \\
             &  & \ion{Zn}{ii} & 37 & 0.29 & 2.5 \\
                             &  & $z<$2.34   & 37 & 0.39 & 3.3 \\
                              & & $z>$2.34   & 37 & 0.34 & 2.9 \\
\hline
${\rm [Z/H]}$ & $v_{\rm max, \cf}$ & all & 74 & 0.24  &   2.9 \\
              &  & interv DLA & 58  & 0.24 & 2.6 \\
             &  & \ion{Zn}{ii} & 37 & 0.22 & 1.9 \\
\hline
                            $N_{\rm \cf}$ & $l_{\rm c}$ & \cts\ detected & 21 & 0.30 & 1.9 \\
\hline
\end{tabular}
\end{minipage}
\end{table}

%% file: tab3.tex
\begin{table}
\begin{minipage}[t]{\columnwidth}
\caption{DLAs vs sub-DLAs, and intervening vs proximate systems}
\begin{tabular}{lllll c}
\hline\hline
Category\footnote{Each entry in this table shows the mean and standard deviation of the given property in the given category.} & \# & $\langle$log\,$N_{\rm \cf}\rangle$& $\langle\Delta v_{\rm \cf}\rangle$ & $\langle|\bar{v}_{\rm \cf}|\rangle$ & $\langle{\rm log}\,N_{\rm \hw}\rangle$ \\
 & & ($N$ in \sqcm) & (km\,s$^{-1}$) & (km\,s$^{-1}$) & ($N$ in \sqcm)\\
\hline
     DLAs & 63 & 14.27$\pm$0.57 & 251$\pm$211 &  68$\pm$83 & 19.77$\pm$0.44 \\
 Sub-DLAs & 11 & 14.07$\pm$0.67 & 272$\pm$281 &  73$\pm$54 & 19.33$\pm$0.48 \\
\hline
    Intervening & 67 & 14.28$\pm$0.58 & 264$\pm$226 &  71$\pm$81 & 19.73$\pm$0.47 \\
                                                                               Proximate\footnote{Proximate absorbers are those within 5\,000\,\kms\ of the QSO redshift. All others are intervening} &  7 & 13.93$\pm$0.57 & 153$\pm$131 &  43$\pm$58 & 19.48$\pm$0.38 \\
\hline
\end{tabular}
\end{minipage}
\end{table}